\documentclass[prd,twocolumn,showpacs,showkeys,footinbib]{revtex4}
\usepackage{amsmath}
\usepackage{graphicx}
\usepackage{slashed}

\renewcommand\vec[1]{{\boldsymbol{\bf #1}}}
\newcommand\adj[1]{\bar{#1}}
\newcommand\he[1]{#1^{\dagger}}
\DeclareMathOperator{\tr}{Tr}
\newcommand\D{\mathcal D}
\newcommand\G{\mathcal G}
\newcommand\ef{\epsilon_{\text{F}}}
\newcommand\kf{k_{\text{F}}}
\newcommand\sumint{\hbox{$\sum$}\!\!\!\!\!\!\!\int}

\begin{document}

\title{Strongly interacting Fermi systems in $1/N$ expansion:\\
       From cold atoms to color superconductivity}
\author{Hiroaki Abuki}
\email{abuki@th.physik.uni-frankfurt.de}
\affiliation{Institut f\"ur Theoretische Physik, Goethe-Universit\"at,
Max-von-Laue-Stra\ss e 1, D-60438 Frankfurt am Main, Germany}
\author{Tom\'{a}\v{s} Brauner\footnote{On leave from Department of Theoretical
Physics, Nuclear Physics Institute ASCR, CZ-25068 \v Re\v z, Czech Republic}}
\email{brauner@ujf.cas.cz}
\affiliation{Institut f\"ur Theoretische Physik, Goethe-Universit\"at,
Max-von-Laue-Stra\ss e 1, D-60438 Frankfurt am Main, Germany}

\begin{abstract}
We investigate the $1/N$ expansion proposed recently as a strategy to include
quantum fluctuation effects in the nonrelativistic, attractive Fermi gas at and
near unitarity. We extend the previous results by calculating the
next-to-leading order corrections to the critical temperature along the
whole BCS--BEC crossover. We demonstrate explicitly that the extrapolation from
the mean-field approximation, based on the $1/N$ expansion, provides a useful
approximation scheme only on the BCS side of the crossover. We then apply the
technique to the study of strongly interacting relativistic many-fermion
systems. Having in mind the application to color superconductivity in cold dense
quark matter, we develop, within a simple model, a formalism suitable to
compare the effects of order parameter fluctuations in phases with
different pairing patterns. Our main conclusion is that the relative
correction to the critical temperature is to a good accuracy proportional to
the mean-field ratio of the critical temperature and the chemical potential. As
a consequence, it is significant even rather deep in the BCS regime, where
phenomenologically interesting values of the quark--quark coupling are
expected. Possible impact on the phase diagram of color-superconducting quark
matter is discussed.
\end{abstract}

\maketitle

\section{Introduction}
Strongly interacting many-fermion systems have been a theoretical challenge for
decades. While the Bardeen--Cooper--Schrieffer (BCS), or mean-field (MF),
theory provides an excellent description of conventional metallic
superconductors, it is still inappropriate for a large range of other systems,
from high-temperature superconductors to nuclear (and more recently, quark)
matter.

Great theoretical interest was triggered by the conjecture that when the
interaction strength is increased, BCS-type superconductivity evolves smoothly
to the Bose--Einstein condensation (BEC) of tightly bound difermion molecules
\cite{Eagles:1969ea,Leggett:1980le,Nozieres:1985no}. The crucial observation in
this respect was that the standard BCS superconducting ground state has the
same form as the ground state of a condensed Bose gas once the composite
operator creating a Cooper pair is identified with that of a bosonic
quasiparticle. However, in spite of the successful unified description of BCS
superconductivity and BEC, a quantitative understanding of the crossover
between the two regimes was missing.

Spectacular progress in this direction has been made in the past decade thanks
to the experiments using ultracold atomic Fermi gases
\cite{Regal:2004re,Bartenstein:2004ba,Zwierlein:2004zw,Bourdel:2004bo}. From
the theorists' point of view, these provide an ideal tool to test the
developing many-body techniques. In particular, a lot of interest has been
attracted by the transition regime between the BCS and BEC limits---the unitary
Fermi gas. In this case, the two-body scattering length is much larger (ideally
infinite) than any other characteristic scale of the system such as the
interaction range and interparticle distance. This, on the one hand, leads to
intriguing universal behavior with connections to other branches of physics
such as the quark--gluon plasma or even string theory
\cite{Son:2008ye,Balasubramanian:2008dm}. On the other hand, it poses a
challenging problem due to the lack of a small expansion parameter.

Several approaches have been suggested to deal with the Fermi gas at large
scattering length, including various self-consistent resummation techniques for
the many-body Green's functions \cite{Haussmann:1994ha,Haussmann:2007ha} or the
scattering matrix \cite{Heiselberg:2001he}, expansion in the dimensionality of
the space \cite{Nussinov:2006nu,Nishida:2006br,Chen:2006wx}, or the $1/N$
expansion \cite{Nikolic:2007ni,Veillette:2007ve}. An early review of the
many-body approaches to the crossover problem may be found in Ref.
\cite{Chen:2005ch}. The predictions of the analytic approximation schemes as
well as the available experimental results are now being tested by increasingly
precise numerical simulations
\cite{Carlson:2003ca,Bulgac:2005pj,Burovski:2006bu,Burovski:2008bu}.

Also in high-energy physics has the mechanism of Cooper pairing proven
extremely fruitful. The Nambu--Jona-Lasinio (NJL) model
\cite{Nambu:1961tp,Nambu:1961fr}, constructed in direct analogy with the BCS
theory of superconductivity, was one of the first models of dynamical symmetry
breaking. For its simplicity and generality, it still remains a popular
low-energy effective description of strongly interacting quark and nuclear
matter, this article not being an exception.

While the original NJL model dealt with dynamical breaking of chiral symmetry
by particle--antiparticle correlations, the true Cooper pairing of two
relativistic fermions near their Fermi surface appears in dense quark matter,
leading to the so-called color superconductivity
\cite{Barrois:1977xd,Frautschi:1978rz} (see \cite{Alford:2007rw} for a recent
review). Due to the strong-interaction nature of quantum chromodynamics (QCD),
quark matter at moderate densities is a typical example of a relativistic
many-fermion system where a departure from the BCS-like behavior is to be
expected. The early work in this respect focused on the structural change
of Cooper pairs at strong coupling and precursor phenomena above the critical
temperature for the superconducting phase transition, in particular the
appearance of the pseudogap in the spectrum
\cite{Babaev:1999iz,Matsuzaki:1999ww,Abuki:2001be,Kitazawa:2003cs,Nawa:2005sb,
Kitazawa:2007zs}.

In principle, the size of the QCD running coupling depends just on the
energy/momentum scale and is thus fixed by the density of the quark matter.
However, in order to better understand the strong-coupling effects in
color superconductors, one often considers models with variable coupling
strength where the full BCS--BEC crossover can be studied
\cite{Nishida:2005ds,Abuki:2006dv,Deng:2006ed,He:2007kd,Sun:2007fc,Brauner:2008td,Chatterjee:2008dr}.
In the color superconductivity context, most of the calculations are still done
using the MF approximation. The first attempts to include the fluctuations of
the order parameter have adapted the Nozi\`eres--Schmitt-Rink (NSR) theory
\cite{Nozieres:1985no,Nishida:2005ds,Abuki:2006dv} and the pseudogap
approximation \cite{He:2007yj}, commonly used in condensed-matter physics, and
the Cornwall--Jackiw--Tomboulis formalism \cite{Cornwall:1974vz,Deng:2008ah},
well known in high-energy physics.

The goal of this paper is to investigate the applicability of the $1/N$
expansion to the study of strongly interacting Fermi gases. In Sec.
\ref{Sec:NRgas} we review the $1/N$ expansion for the nonrelativistic Fermi gas
developed in Refs. \cite{Nikolic:2007ni,Veillette:2007ve}. We study the
evolution of the next-to-leading-order corrections from the unitarity towards
the BCS and BEC regimes and critically examine the virtues as well as
shortcomings of the method. In Sec. \ref{Sec:rel_matter} we then apply the
technique to relativistic superconductors, using a class of NJL-type models. In
particular we estimate the correction to the critical temperature in
color-superconducting quark matter and show that it is significant even for
realistic values of the coupling strength. On account of the fact that the
fluctuation effects are different for various competing phases, we propose a
modification of the QCD phase diagram. Finally, in Sec. \ref{Sec:conclusions} we
summarize and conclude.

\section{Nonrelativistic attractive Fermi gas}
\label{Sec:NRgas} We consider here the gas of two nonrelativistic fermion
species (``flavors'') which will be referred to as $\psi_\uparrow$ and
$\psi_\downarrow$. For simplicity they will be assumed to have equal masses and
chemical potentials (and thus also densities). Nevertheless, the results of the
present analysis may be straightforwardly generalized to the case of a density
imbalance, or, Fermi surface mismatch \cite{Veillette:2007ve}. At low density,
i.e., low characteristic momentum set by the Fermi scale, the two-body
interaction is completely determined by the $s$-wave scattering length, $a$.

\subsection{Formalism}
\label{Subsec:NRformalism} In order to employ the $1/N$ expansion, one has to
generalize the system by including $N$ copies of the two fermion flavors.
Following Refs. \cite{Nikolic:2007ni,Veillette:2007ve}, we write down the
Euclidean Lagrangian in the form
\begin{multline}
\mathcal L=\sum_{i=1}^N\sum_{\sigma=\uparrow,\downarrow}\he\psi_{i\sigma}
\left(\partial_\tau-\frac{\nabla^2}{2m}-\mu\right)\psi_{i\sigma}\\
-\frac gN\sum_{i,j=1}^N\he\psi_{i\uparrow}\he\psi_{i\downarrow}
\psi_{j\downarrow}\psi_{j\uparrow} .
\label{NRlagrangian}
\end{multline}
The sums over repeated indices will from now on be implicitly assumed. For
$N=1$ the Lagrangian reduces to one describing a two-flavor gas with local
contact attractive interaction with strength $g$ \cite{Sa:1993sa}. In this
extended version the coupling is rescaled as $g\to g/N$ in order to ensure that
the action scales naturally with $N$.

Note that the Lagrangian \eqref{NRlagrangian} possesses, apart from the phase
invariance generated by the total particle number, a symplectic symmetry,
$\mathrm{Sp}(2N)$. However, as will become clear soon, this symmetry
remains unbroken by the Cooper pairing so that it does not give rise to any
unwanted Nambu--Goldstone (NG) bosons and hence does not affect the
low-temperature thermodynamics of the system.

As a next step the theory is bosonized by introducing the auxiliary field,
$\phi\sim\frac gN\psi_{i\downarrow}\psi_{i\uparrow}$, and performing the
Hubbard--Stratonovich transformation. The result is the nonlocal effective
action,
\begin{equation}
\mathcal S=N\int_0^\beta d\tau\int d^3\vec x
\frac{|\phi(\vec x,\tau)|^2}g-N\tr\log\G^{-1}[\phi(\vec x,\tau)],
\label{NRaction}
\end{equation}
where $\G$ is the Nambu-space fermion propagator in presence of the pairing
field $\phi$,
$$
\G^{-1}=\begin{pmatrix}
-\partial_\tau+\frac{\nabla^2}{2m}+\mu & \phi\\
\phi^* & -\partial_\tau-\frac{\nabla^2}{2m}-\mu
\end{pmatrix}.
$$
Eq. \eqref{NRaction} can be interpreted as a classical action that defines a
theory of the scalar field $\phi$ with self-interactions determined by the
expansion of the action in powers of $\phi$. The crucial observation made in
Refs. \cite{Nikolic:2007ni,Veillette:2007ve} is that since the action is
proportional to $N$, the expansion of the partition function, or
the thermodynamic potential, in powers of $1/N$ is equivalent to the expansion
in loops. Here and in the following, the term ``loop'' is used to refer to a
bosonic loop, unless explicitly indicated otherwise. Note that all loops
containing fermions of the original theory \eqref{NRlagrangian} are resummed
into the action \eqref{NRaction}, i.e., are included at the tree level with
respect to bosons.

Obviously, the leading order (LO) of the $1/N$ expansion is equivalent to the
saddle-point approximation to the functional integral, that is, the usual MF
approximation. The next-to-leading order (NLO) then incorporates one-loop
corrections, or, the Gaussian fluctuations around the saddle point
\cite{Diener:2007di}. The $1/N$ expansion thus provides a systematic ordering
of the corrections to the MF approximation. We should nevertheless keep in mind
that at the end of the calculation, we have to set $N=1$. The way this
extrapolation is performed is to be understood as a part of the definition of
the method, which distinguishes it from other approaches with formally
equivalent thermodynamic potential \cite{Nozieres:1985no,Diener:2007di}.

In general one calculates, in a given approximation scheme, the thermodynamic
potential $\Omega$ as a function of the anticipated vacuum expectation value
$\Delta$ of the field $\phi$, and of the chemical potential $\mu$. Their actual
values in thermodynamic equilibrium are then determined by a simultaneous
solution of the gap and number equations,
\begin{equation}
\frac{\partial\Omega}{\partial\Delta}=0,\quad
\frac{\partial\Omega}{\partial\mu}=-n,
\label{gap_num_eqs}
\end{equation}
where $n$ is the total particle density, related to the Fermi momentum $\kf$ by
the usual expression, $n=\kf^3/3\pi^2$. It is well known that already at
one-loop level, an attempt to solve the equations self-consistently leads to
unphysical results, in particular the violation of the Goldstone theorem
\cite{Haussmann:2007ha}. Veillette \emph{et al.} \cite{Veillette:2007ve}
suggested to avoid this problem by an expansion of the gap and chemical
potential simultaneously with the expansion of the thermodynamic
potential that follows from the action \eqref{NRaction},
\begin{align*}
\Omega&=N\Omega^{(0)}+\Omega^{(1)}+\frac1N\Omega^{(2)}+\dotsb,\\
\Delta&=\Delta^{(0)}+\frac1N\Delta^{(1)}+\frac1{N^2}\Delta^{(2)}+\dotsb,\\
\mu&=\mu^{(0)}+\frac1N\mu^{(1)}+\frac1{N^2}\mu^{(2)}+\dotsb.
\end{align*}
Comparing terms of the same order in the gap and number equations
\eqref{gap_num_eqs}, one obtains \emph{explicit} expressions for the
higher-order corrections to the MF values $\Delta^{(0)},\mu^{(0)}$. In
particular at NLO we find
\begin{equation}
\begin{pmatrix}
\mu^{(1)}\\ \Delta^{(1)}
\end{pmatrix}
=-
\begin{pmatrix}
\partial_{\mu\mu}\Omega^{(0)} & \partial_{\mu\Delta}\Omega^{(0)}\\
\partial_{\Delta\mu}\Omega^{(0)} & \partial_{\Delta\Delta}\Omega^{(0)}
\end{pmatrix}^{-1}
\begin{pmatrix}
\partial_\mu\Omega^{(1)}\\ \partial_\Delta\Omega^{(1)}
\end{pmatrix}.
\label{NLO_corr_general}
\end{equation}
It is essential that all derivatives of the thermodynamic potential here are
evaluated using the MF values $\Delta^{(0)},\mu^{(0)}$. One thus avoids the
problems with self-consistency; in particular the NG boson of
the spontaneously broken symmetry is exactly massless \cite{Nikolic:2007ni}.

When we merely wish to determine the critical temperature, the gap in Eq.
\eqref{gap_num_eqs} is fixed to zero and we solve for the temperature
and chemical potential instead. The identity \eqref{NLO_corr_general} then
naturally modifies to one for the corrections of the variables of interest,
\begin{equation}
\begin{pmatrix}
\mu_c^{(1)}\\ T_c^{(1)}
\end{pmatrix}
=-
\begin{pmatrix}
\partial_{\mu\mu}\Omega^{(0)} & \partial_{\mu T}\Omega^{(0)}\\
\partial_{\Delta\mu}\Omega^{(0)} & \partial_{\Delta T}\Omega^{(0)}
\end{pmatrix}^{-1}
\begin{pmatrix}
\partial_\mu\Omega^{(1)}\\ \partial_\Delta\Omega^{(1)}
\end{pmatrix}.
\label{NLO_corr_Tc}
\end{equation}
We will comment later on the ambiguity that arises at this point, stemming from
the fact that we can choose to solve Eq. \eqref{gap_num_eqs} for $1/T_c$ (or
any other function of $T_c$) and accordingly get a different type of $1/N$
expansion---all derivatives with respect to $T$ in Eq. \eqref{NLO_corr_Tc}
would simply turn into ones with respect to $1/T$.

Eqs. \eqref{NLO_corr_general} and \eqref{NLO_corr_Tc} of course have to be
supplemented with an expression for the thermodynamic potential. For the sake
of this section, we will need just the explicit form of the standard MF
part,
\begin{multline}
\Omega^{(0)}=\frac{|\Delta|^2}g
-\int\frac{d^3\vec k}{(2\pi)^3}(E_{\vec k}-\xi_{\vec k})\\
-2T\int\frac{d^3\vec k}{(2\pi)^3}\log\left(1+e^{-\beta E_{\vec k}}\right),
\label{NRf0}
\end{multline}
using the usual notation $\epsilon_{\vec k}=\frac{\vec k^2}{2m}$, $\xi_{\vec
k}=\epsilon_{\vec k}-\mu$, and $E_{\vec k}=\sqrt{\xi_{\vec k}^2+|\Delta|^2}$.
The
bare coupling $g$ is related to the physical scattering length by
\begin{equation}
\frac1g=-\frac{m}{4\pi a}+\int\frac{d^3\vec k}{(2\pi)^3}\frac1{2\epsilon_{\vec
k}}.
\label{renorm_prescr}
\end{equation}
For further details, we refer the reader to the original literature
\cite{Nikolic:2007ni,Veillette:2007ve} where all necessary formulas are derived
in detail. In Sec. \ref{Sec:rel_matter} we will develop a relativistic
formalism from which the present case will follow as a particular
nonrelativistic limit.

\subsection{Numerical results}
\label{Subsec:NRresults}
\subsubsection{Critical temperature}
While differing in other specific directions of investigation, both Refs.
\cite{Nikolic:2007ni,Veillette:2007ve} addressed the question of special
interest, the calculation of the critical temperature at unitarity. Veillette
\emph{et al.} obtained the result
\begin{equation}
\frac{T_c}{\ef}=0.4964-\frac{1.31}N
\label{NLO_Tc_Veillette}
\end{equation}
in units of the Fermi energy, $\ef=\kf^2/2m$, whereas Nikoli\'c and Sachdev
calculated the correction to the inverse temperature and got
\begin{equation}
\frac{\ef}{T_c}=2.014+\frac{5.317}N.
\label{NLO_Tc_Nikolic}
\end{equation}
Both results are formally equivalent to order $1/N$ in the expansion, yet they
yield dramatically different numbers when evaluated at $N=1$. Indeed, Eq.
\eqref{NLO_Tc_Veillette} becomes even meaninglessly negative and can be merely
used to make the qualitative conclusion that the fluctuations decrease the
critical temperature significantly.

On the other hand, Eq. \eqref{NLO_Tc_Nikolic} leads to the critical temperature
$T_c=0.14\ef$ in remarkable agreement with the result $T_c=0.152(7)\ef$,
obtained by numerical Monte Carlo simulations
\cite{Burovski:2006bu,Burovski:2008bu}. However, it should be stressed that
there is no \emph{a priori} criterion that
would tell us which observable to choose for the evaluation of the NLO
correction. One may think that it is the Lagrange multiplier $\beta=1/T$ rather
than the temperature itself that is the natural variable of the thermodynamic
potential. Still a deeper physical argument is obviously needed to resolve this
ambiguity. Here we just remark that on the technical level, it clearly arises
from the truncation of the $1/N$ series for different observables; one cannot
expect the $1/N$ expansion to be reliable when the NLO term is larger than the
LO one \footnote{It was argued in Ref. \cite{Nikolic:2007ni} that due to the
distinct physical origin of the LO (fermions) and NLO (bosons) contribution to
the thermodynamic potential, their relative size should not be used to
test the accuracy of the $1/N$ expansion. This postpones the issue of
convergence of the series to higher orders, yet the ambiguity encountered at
NLO remains.}.

\begin{figure}
\includegraphics[scale=1.1]{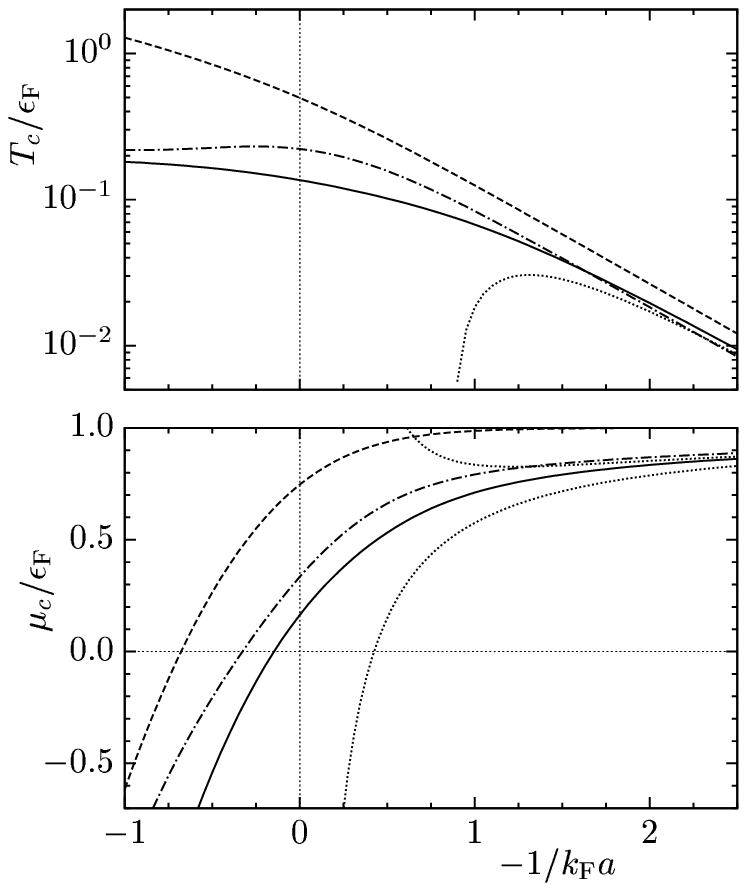}
\caption{Critical temperature and chemical potential as a function of the
inverse scattering length. Dashed lines: MF approximation (LO in $1/N$). Solid
lines: NLO calculation based on expansion of $1/T$. Dotted line in the upper
panel: NLO value based on expansion of $T$ as in Eq. \eqref{NLO_corr_Tc}.
Dash-dotted lines: Self-consistent calculation using NSR theory. Dotted lines
in the lower panel: First- and second-order perturbative approximations to the
chemical potential, see Eq. \eqref{PT_mu}.}
\label{Fig:NR_Tc}
\end{figure}

In order to further study the size of the $1/N$ corrections and the sensitivity
to the choice of observable to evaluate them, we calculated the critical
temperature at NLO as a function of the inverse scattering length, see Fig.
\ref{Fig:NR_Tc}.

Obviously, the evaluation of the critical temperature based on Eq.
\eqref{NLO_corr_Tc} (dotted line in the upper panel) is only reasonable in the
far BCS regime (right end of the plot) where it roughly coincides with the
$1/T$-based approximation (solid line). The $1/T$-based value will therefore be
used exclusively in the following and will be referred to simply as the NLO
critical temperature. For comparison we also show the critical temperature and
chemical potential calculated using the NSR theory (dash-dotted line). NSR
theory takes the fluctuation effects into account only in the total particle
number, but not in the gap equation. On the other hand, it then solves
Eq. \eqref{gap_num_eqs} self-consistently \footnote{Strictly speaking, our
calculation differs from the original NSR theory \cite{Nozieres:1985no} in that
we perform the Matsubara sum in the one-loop contribution to the number density
directly while it is more conventional to make analytic continuation to the real
frequencies and use spectral representation. Of course, the results should
coincide as long as the analytic continuation has been done properly.}.

In the BCS regime the critical temperatures calculated within the two
approaches are consistent with each other, at least qualitatively. Around
unitarity, NSR theory predicts a well-known tiny maximum. On the contrary,
the $1/N$ expansion gives a monotonic dependence on the inverse scattering
length, with the value at unitarity agreeing very well with Monte Carlo
calculations \cite{Burovski:2006bu,Burovski:2008bu}, as noted above. However, as
also emphasized there, the $1/N$ expansion should not be really trusted in this
region without an additional physical insight. Finally, in the BEC limit NSR
theory converges to the expected critical temperature of the free Bose gas,
$T_c=0.218\ef$. $1/N$ expansion fails to reproduce this asymptotic behavior.
Although it cannot be seen in Fig. \ref{Fig:NR_Tc}, its critical temperature
keeps growing towards the BEC limit, albeit with a decreasing slope. Indeed, one
cannot really expect to get a constant asymptotics from a perturbative expansion
around the rapidly increasing MF value.

In the lower panel of Fig. \ref{Fig:NR_Tc} we also display the results for the
chemical potential at the critical temperature. Here the most interesting is
the evolution in the far BCS region. While in the MF approximation the chemical
potential approaches its asymptotic value equal to the Fermi energy with an
exponentially decreasing tail, upon including fluctuations the convergence turns
into a much slower, power-law one. This is a purely perturbative effect, as
clearly demonstrated by a comparison with the standard perturbative expansion
for the chemical potential in the dilute Fermi gas \cite{Fetter:1971fw} (the
first- and second-order values are plotted using dotted lines)
\begin{equation}
\frac{\mu}{\ef}=1+\frac4{3\pi}\kf a+\frac{4(11-2\log2)}{15\pi^2}(\kf a)^2+\dotsb.
\label{PT_mu}
\end{equation}
Even though this formula holds only in the normal phase and at zero
temperature, the agreement is excellent. The reason, of course, is that the
pairing effects are suppressed exponentially and finite-temperature effects are
also suppressed by the exponentially small value of the critical temperature in
the BCS limit, so that they are both completely negligible with respect to the
perturbative corrections in Eq. \eqref{PT_mu} \cite{Diener:2007di}.

\begin{figure}
\includegraphics[scale=0.8]{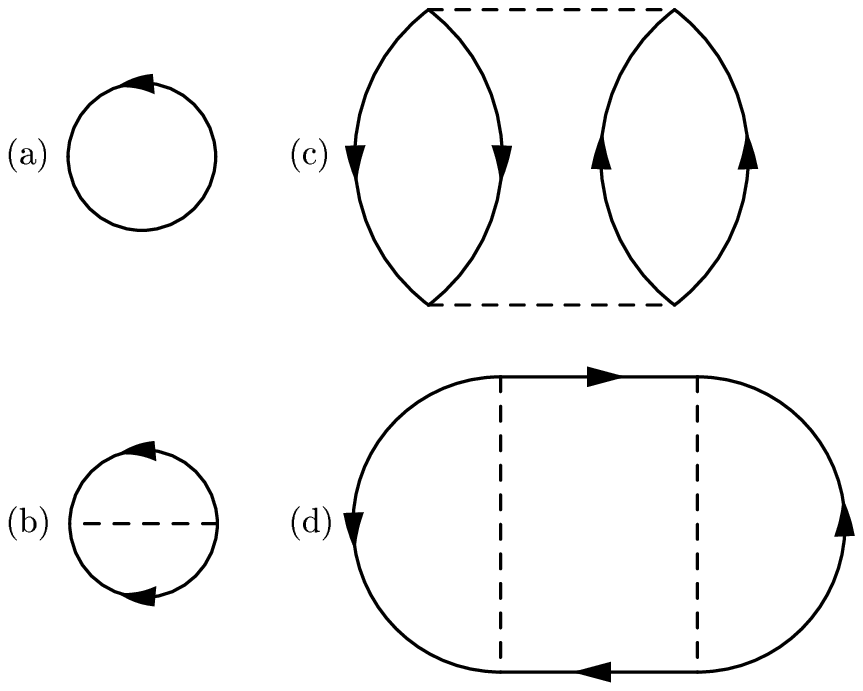}
\caption{Lowest-order perturbative contributions to the thermodynamic potential
of the Fermi gas. The free-gas graph (a) coincides with LO in $1/N$. The
first-order graph (b) and the second-order graph (c) appear at NLO in $1/N$.
The second-order graph (d) contributes only at NNLO. The dashed line denotes
the bare propagator of the pairing field $\phi$, $g/N$.}
\label{Fig:PTgraphs}
\end{figure}

The fact that the NLO in $1/N$ reproduces the perturbative expansion of the
chemical potential up to second order can also be verified directly on the
Feynman graph level. In Fig. \ref{Fig:PTgraphs} we show the perturbative
contributions to the thermodynamic potential. In general the $1/N$ expansion
does not coincide with perturbation series---the diagram (d) is of second order
in the interaction, yet only appears at the next-to-next-to-leading order in
$1/N$. Fortunately, it turns out to vanish at zero temperature so that the $1/N$
expansion to NLO indeed contains the full second-order perturbative correction
\cite{Furnstahl:2002gt}.

\subsubsection{BCS limit}
Since we have concluded that the $1/N$ expansion is most reliable on the BCS
side of the crossover, let us now investigate this regime in more detail. The
matrix of second derivatives of $\Omega^{(0)}$ at the critical temperature
needed to evaluate the NLO corrections $\beta^{(1)}$ and $\mu^{(1)}$ can be
easily calculated from Eq. \eqref{NRf0}. One finds
\begin{gather*}
\partial_{\mu\mu}\Omega^{(0)}=-\frac{3n}{2\ef},\quad
\partial_{\mu\beta}\Omega^{(0)}=\frac{\pi^2nT^3}{4\ef^2},\\
\partial_{\Delta^2\beta}\Omega^{(0)}=-\frac{3nT}{4\ef},
\end{gather*}
up to corrections which are exponentially small as $\kf a\to0-$. (Note that we
prefer to take the derivative with respect to $\Delta^2$ instead of $\Delta$
since at the critical temperature we have to set $\Delta=0$ afterwards.) The
last needed coefficient,
$$
\partial_{\Delta^2\mu}\Omega^{(0)}=\frac{3n}{4\ef^2}\left(\frac{\pi}{4\kf
a}-1\right),
$$
is most easily obtained using the exact relation
$$
\partial_{\Delta^2\beta}\Omega^{(0)}=-\frac{3\pi nT}{16\ef\kf a}+\mu
T\partial_{\Delta^2\mu}\Omega^{(0)},
$$
which follows from the observation that $\Omega^{(0)}$ can be expressed in
terms of a dimensionless function of the combination $\beta\mu$.

The coefficient $\partial_{\mu\beta}\Omega^{(0)}$ is strongly suppressed by the
third
power of the critical temperature so that it may be neglected and the required
matrix of second derivatives [as in Eq. \eqref{NLO_corr_Tc}, just with modified
variables] becomes
\begin{multline*}
\begin{pmatrix}
\partial_{\mu\mu}\Omega^{(0)} & \partial_{\mu\beta}\Omega^{(0)}\\
\partial_{\Delta^2\mu}\Omega^{(0)} & \partial_{\Delta^2\beta}\Omega^{(0)}
\end{pmatrix}^{-1}\\
\approx
\frac1{\partial_{\mu\mu}\Omega^{(0)}\partial_{\Delta^2\beta}\Omega^{(0)}}
\begin{pmatrix}
\partial_{\Delta^2\beta}\Omega^{(0)} & 0\\
-\partial_{\Delta^2\mu}\Omega^{(0)} & \partial_{\mu\mu}\Omega^{(0)}
\end{pmatrix}.
\end{multline*}
As a result, the leading NLO contribution to the chemical potential decouples
and is solely determined by $\partial_\mu\Omega^{(1)}$ and
$\partial_{\mu\mu}\Omega^{(0)}$,
$$
\mu_c^{(1)}=-\frac{\partial_\mu\Omega^{(1)}}{\partial_{\mu\mu}\Omega^{(0)}}.
$$
This is not surprising since we know from Eq. \eqref{PT_mu} that in the BCS
limit the chemical potential is governed by perturbative effects. The expression
for the shift of the inverse critical temperature also simplifies to
$$
\beta_c^{(1)}=-\frac{\partial_{\Delta^2}\Omega^{(1)}+\mu_c^{(1)}\partial_{
\Delta^2\mu } \Omega^ {(0)}}{\partial_{\Delta^2\beta}\Omega^{(0)}}.
$$
Substituting all the analytic expressions listed above as well as the chemical
potential correction from Eq. \eqref{PT_mu} we get the final prediction of the
$1/N$ expansion for the NLO relative shift of the inverse temperature in the BCS
limit,
$$
\frac{\beta_c^{(1)}}{\beta_c^{(0)}}=\frac{4\ef}{3n}\partial_{\Delta^2}\Omega^{
(1)}
+\frac13\left(1-\frac{9+2\log2}{5\pi}\kf a\right),
$$
where the only missing ingredient, that has to be evaluated numerically, is
$\partial_{\Delta^2}\Omega^{(1)}$.

The point of these considerations is that the slow, perturbative convergence of
the chemical potential results in a rather large offset in the critical
temperature, which \emph{survives} even in the limit $\kf a\to0-$. This is a
consequence of the fact that $\partial_{\Delta^2\mu}\Omega^{(0)}$ diverges in
the BCS limit; we regard it an artifact of the expansion of the gap
and number equations, leading to the expression \eqref{NLO_corr_general}.

\begin{figure}
\includegraphics[scale=1.1]{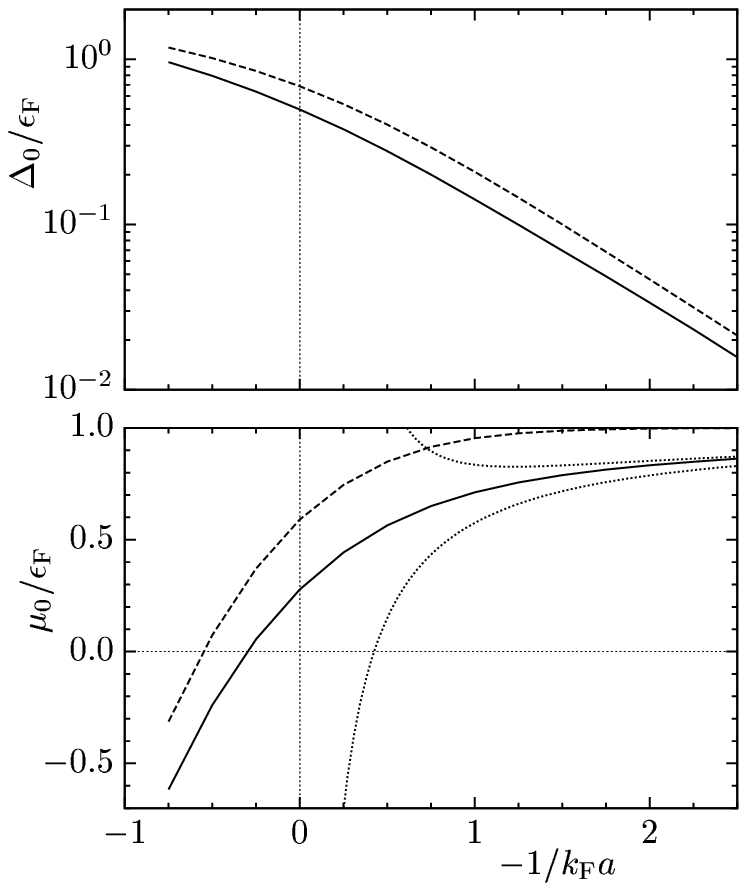}
\caption{Gap and chemical potential at zero temperature as a function of the
inverse scattering length. Dashed lines: MF approximation (LO in $1/N$). Solid
lines: NLO calculation based on expansion of $\Delta^2$. Dotted lines in lower
panel: First- and second-order perturbative approximations to the chemical
potential, see Eq. \eqref{PT_mu}.}
\label{Fig:NR_T0}
\end{figure}
In addition to the critical temperature, we have also computed the gap and
chemical potential at zero temperature, see Fig. \ref{Fig:NR_T0}. This has
already been done by Veillette \emph{et al.} \cite{Veillette:2007ve}, although
they did not investigate the asymptotic behavior in the BCS limit. Our
calculation differs from theirs only in that we have for technical reasons taken
$\Delta^2$ instead of $\Delta$ as the variable to make the $1/N$ expansion.
Since the relative NLO correction to the gap at zero temperature is small, the
effect of this change is nearly negligible.

\begin{figure}
\includegraphics[scale=1.1]{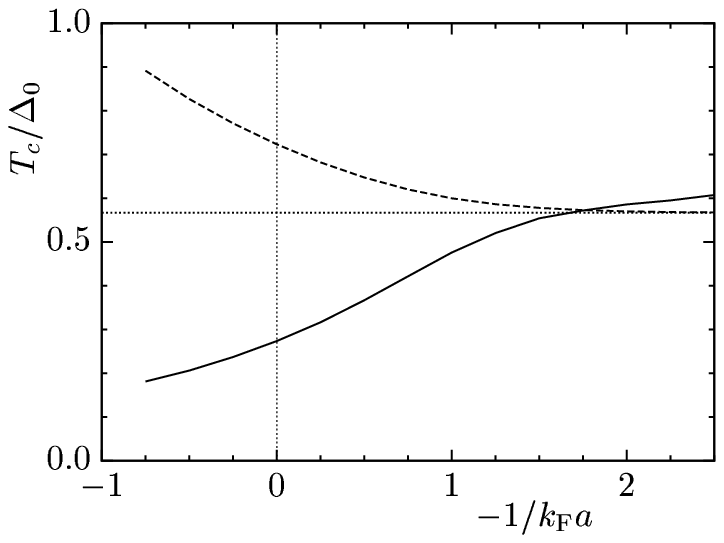}
\caption{Ratio of the critical temperature to the gap at zero temperature as a
function of the inverse scattering length. Dashed line: MF approximation. Solid
line: NLO calculation with data taken from Figs. \ref{Fig:NR_Tc} and
\ref{Fig:NR_T0}. Horizontal dotted line: Prediction of the BCS theory.}
\label{Fig:BCS_ratio}
\end{figure}
Fig. \ref{Fig:NR_T0} suggests that the gap also acquires a constant offset
that survives the BCS limit (even though we were not able to check this
conclusion analytically). However, the offset is not the same as in the case
of the critical temperature, as is clearly seen from Fig. \ref{Fig:BCS_ratio}
where we plot the ratio $T_c/\Delta_0$. Thus, $1/N$ expansion predicts a
departure of this ratio from the BCS value $e^\gamma/\pi$. In Sec.
\ref{Sec:rel_matter} we will present a calculation of the critical temperature
within a $1/N$-inspired high-density approximation, where such artifacts will
be absent.

\section{Dense relativistic matter}
\label{Sec:rel_matter}
We are now going to apply the $1/N$ expansion to pairing in relativistic
systems. Physically, this amounts to including the antiparticles among the
degrees of freedom, and to modifying the fermion dispersion relation. In
particular the change of dispersion relation affects the ultraviolet structure
of the theory, leading to new divergences that are not present in the
nonrelativistic case \cite{Abuki:2006dv}. In Sec. \ref{Subsec:HDET} we evade
this difficulty by making a suitable approximation, appropriate in the far BCS
regime.

\subsection{NJL-type model}
\label{Subsec:NJLmodel}
Following closely the notation of Ref. \cite{Brauner:2008td}, we consider a
class of NJL-type models defined by the Lagrangian
\begin{equation}
\mathcal L=\adj\psi(i\slashed\partial+\mu\gamma_0-m)\psi+\frac g4\sum_a
|\adj{\psi^{\mathcal C}}\gamma_5Q_a\psi|^2,
\label{R_Lagrangian}
\end{equation}
where $\psi^{\mathcal C}=C\adj\psi^T$ is the standard charge-conjugated Dirac
spinor and the set of matrices $Q_a$, acting on the internal degrees of freedom,
are normalized by $\tr(Q_a\he Q_b)=\delta_{ab}$. Simplified as much as
possible, this Lagrangian describes a system of interacting fermions with equal
masses and chemical potentials. The pairing is assumed to occur in a spin-zero,
positive-parity channel, but its flavor structure, determined by the matrices
$Q_a$, can be arbitrary. Once we understand in detail the fluctuation effects in
this simple setting, we will move on to more realistic systems in our future
work.

\begin{figure}
\includegraphics[scale=0.6]{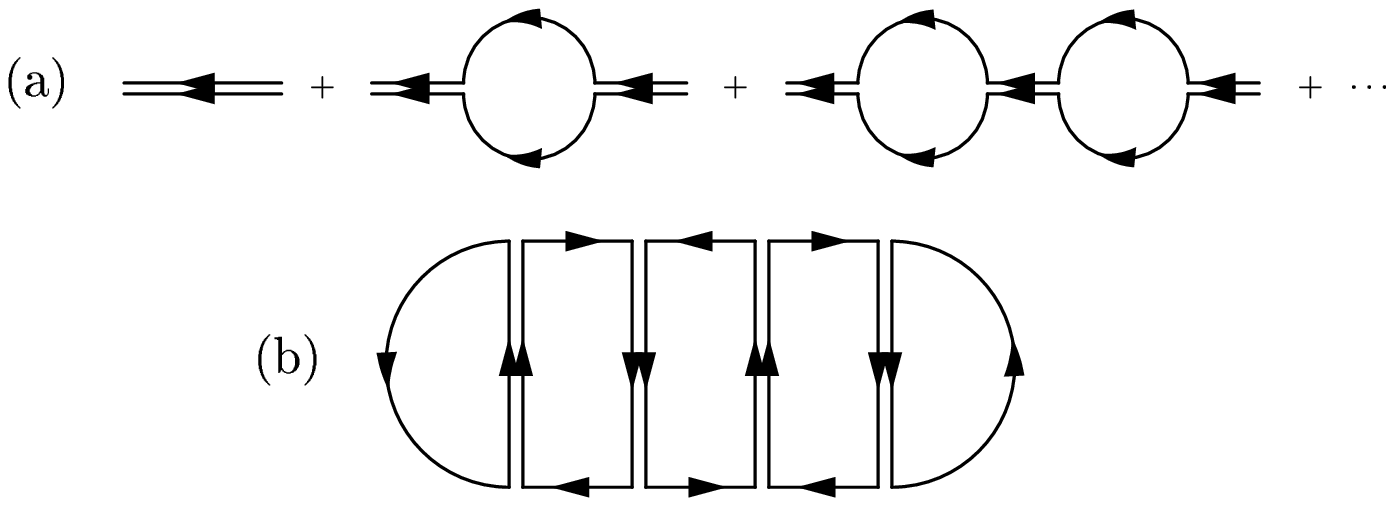}
\caption{(a) RPA propagator of the collective mode in the case it transforms as
an antisymmetric tensor of $\mathrm{SU}(N)$, using the double-line
notation. (b) One of the planar diagrams which dominate the thermodynamic
potential in the large-$N$ limit.}
\label{Fig:ladder}
\end{figure}
The first task to solve is the generalization of the model to arbitrary $N$ so
that we can subsequently make the appropriate expansion. With the application
to quark matter in mind, one may think it would be most convenient to use the
color $\mathrm{SU}(3)$ symmetry already present in the system and extend it to
$\mathrm{SU}(N)$. We would like to explain here in detail why this would not
work.

As a consequence of the QCD interactions, quarks are assumed to pair in a
color-antisymmetric configuration. (Symmetry would not make a difference.
Important is that the Cooper pair cannot be a singlet of the symmetry due to the
complex nature of the group $\mathrm{SU}(3)$.) Upon the generalization of the
theory the pairs would transform in the antisymmetric-tensor representation of
$\mathrm{SU}(N)$. Disregarding the fact that this would lead to a very large
number, of order $\mathcal O(N^2)$, of collective modes, there is another, more
serious problem.

In Sec. \ref{Sec:NRgas} we saw using the scaling of the thermodynamic
potential that the $1/N$ expansion reproduces the MF approximation at the
leading order. This may be also understood in terms of the collective mode
propagator. In the MF approximation, this consists of a geometric series of
graphs in the random-phase approximation (RPA). Now since the boson is a singlet
of the $\mathrm{Sp}(2N)$ symmetry, each fermion bubble contributes a factor $N$
from the trace over the flavor space, thus compensating for factors $1/N$ coming
from the coupling $g$. The point is that when the boson becomes a tensor of
$\mathrm{SU}(N)$ rather than a singlet, the trace factors are lost, as is
most easily visualized with the help of the double-line notation, see Fig.
\ref{Fig:ladder} (a). As a consequence, addition of each new fermion bubble in
the RPA series suppresses the graph by another factor $1/N$ from the coupling so
that the full series will not be resummed at any finite order in $1/N$.

Is it then possible to make the coupling scale just as $\mathcal O(1)$? No,
because the thermodynamic potential is dominated by the planar diagrams with a
single fermion loop (rather than the RPA ones), which contain the same number of
flavor traces as the coupling factors, see Fig. \ref{Fig:ladder} (b). So, in
order that the power of $N$ in Feynman graphs is bounded from above and the
$1/N$ expansion makes sense at all, the coupling has to decrease at least as
$1/N$.

We thus conclude that when the difermion field is defined to be a tensor of the
$\mathrm{SU}(N)$ color group, we will not get the RPA propagator and hence the
pairing instability at any finite order in $1/N$. The $1/N$ expansion in
this case efficiently resums a different class of diagrams than necessary.
To remedy this problem, we introduce a new quantum number to label the fermion
fields, i.e., generalize Eq. \eqref{R_Lagrangian} to
\begin{equation}
\mathcal L=\adj\psi_i(i\slashed\partial+\mu\gamma_0-m)\psi_i+\frac g{4N}\sum_a
|\adj{\psi_i^{\mathcal C}}\gamma_5Q_a\psi_i|^2.
\label{R_Lagrangian_N}
\end{equation}
This has two advantages. First, one does not need to rely on the presence of
the color $\mathrm{SU}(3)$. The construction is absolutely general, regardless
of the actual physical internal degrees of freedom. Second, the Cooper pair is a
singlet with respect to the symmetry transformations acting on the new quantum
number so that there are no additional bosonic degrees of freedom introduced by
the extension of the symmetry and no unwanted NG bosons from its spontaneous
breaking. On the other hand one has to set, just like in the nonrelativistic
case, $N=1$ at the end of the calculation.

With the above argument in mind, we proceed as in Sec. \ref{Sec:NRgas}. Upon
bosonization of the theory \eqref{R_Lagrangian_N} by introducing a set of
auxiliary fields, $\phi_a\sim\frac
g{2N}\adj{\psi_i^{\mathcal C}}\gamma_5Q_a\psi_i$, we arrive at the effective
action
\begin{equation}
\mathcal S=N\int_0^\beta d\tau\int d^3\vec x
\frac{|\phi_a(\vec x,\tau)|^2}g-N\tr\log\G^{-1}[\phi_a(\vec x,\tau)],
\label{Raction}
\end{equation}
with the inverse fermion propagator in the Nambu space given by
$$
\G^{-1}=\begin{pmatrix}
i\slashed\partial+\gamma_0\mu-m & -\phi_a\gamma_5\he Q_a\\
\phi_a^*\gamma_5Q_a & i\slashed\partial-\gamma_0\mu-m
\end{pmatrix}.
$$
From the classical action \eqref{Raction} we can generate the LO (RPA)
propagator of the collective bosonic modes by a second functional derivative.
Within this paper, we will for simplicity restrict our attention to the normal
phase; the extension of the formalism below the critical temperature will be
considered elsewhere. In the normal phase, the LO boson propagator becomes
$\D_{0ab}=\D_0\delta_{ab}$,
\begin{widetext}
\begin{multline}
\frac1N\D_0^{-1}(i\omega_n,\vec p)=\frac1g+\frac12\int\frac{d^3\vec
k}{(2\pi)^3}\left\{\left[1+\frac{m^2+\vec k_+\cdot\vec k_-}
{\epsilon_{\vec k_+}\epsilon_{\vec k_-}}\right]
\left[\frac{f(\epsilon_{\vec k_+}+\mu)+f(\epsilon_{\vec k_-}+\mu)-1}
{i\omega_n+2\mu+\epsilon_{\vec k_+}+\epsilon_{\vec k_-}}+
\frac{1-f(\epsilon_{\vec k_+}-\mu)-f(\epsilon_{\vec k_-}-\mu)}
{i\omega_n+2\mu-\epsilon_{\vec k_+}-\epsilon_{\vec k_-}}\right]\right.\\
\left.+\left[1-\frac{m^2+\vec k_+\cdot\vec k_-}
{\epsilon_{\vec k_+}\epsilon_{\vec k_-}}\right]
\left[\frac{f(\epsilon_{\vec k_+}+\mu)-f(\epsilon_{\vec k_-}-\mu)}
{i\omega_n+2\mu+\epsilon_{\vec k_+}-\epsilon_{\vec k_-}}+
\frac{f(\epsilon_{\vec k_-}+\mu)-f(\epsilon_{\vec k_+}-\mu)}
{i\omega_n+2\mu+\epsilon_{\vec k_-}-\epsilon_{\vec k_+}}\right]\right\},
\label{RLOsusc}
\end{multline}
\end{widetext}
where $\vec k_\pm=\vec k\pm\frac{\vec p}2$ and $f(x)=1/(e^{\beta x}+1)$ is the
Fermi--Dirac distribution function.

Assuming that the order parameter fluctuations do not change the second order
of the phase transition, we can find the critical temperature using the
Thouless criterion \cite{Thouless:1960th}, i.e., by requiring that the
normal-phase boson propagator has a pole (pairing singularity) at zero
(four-)momentum. This is equivalent to the gap equation
$\partial\Omega/\partial\Delta^2=0$ at $\Delta=0$. The expression
$\partial\Omega^{(1)}/\partial\Delta^2$ that appears in the NLO formula for the
critical
point \eqref{NLO_corr_Tc}, is thus seen to represent the one-loop boson
self-energy at zero momentum.

\begin{figure}
\includegraphics[scale=1.2]{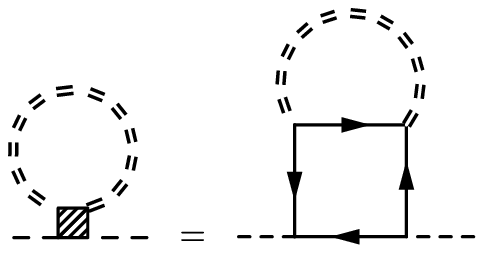}
\caption{One-boson-loop contribution to the collective mode self-energy. The
double-dashed line denotes the resummed LO boson propagator.}
\label{Fig:box_diagram}
\end{figure}
In a general scalar self-interacting theory the one-loop self-energy is given by
the tadpole diagram with one quartic interaction vertex.
In case of the theory defined by the action \eqref{Raction}, the effective
four-boson vertex is generated by a fermion loop, see Fig.
\ref{Fig:box_diagram}. Let us concentrate on the flavor structure of the
diagram. Assuming that the external scalar legs carry the flavor indices $a,b$,
the complete information about the pairing pattern will be encoded in the
flavor trace $\delta_{cd}\tr_{\text F}(Q_a\he Q_b Q_c\he Q_d)$, the Kronecker
delta coming from the internal boson propagator. The combination
$\delta_{cd}Q_c\he Q_d$ is the quadratic Casimir operator of the symmetry group
in the representation of the Cooper pairs, and therefore must be proportional
to the unit matrix as long as this representation is irreducible, i.e., we
consider a single pairing channel \footnote{Should we include several pairing
channels, our results would naturally generalize, giving a different
fluctuation effect for each of them even in the normal phase.}. Writing
$\delta_{cd}Q_c\he Q_d=C_2\openone$ and taking the trace, we immediately find
that $C_2=\delta_{cd}\delta_{cd}/\mathrm{dim}(Q)$, where $\mathrm{dim}(Q)$
denotes the size of the matrices $Q_a$. To conclude the argument we
just note that this is equal to the number of internal \emph{fermionic} degrees
of freedom, $N_{\text F}$, while $\delta_{cd}\delta_{cd}$ counts the number of
\emph{bosonic} degrees of freedom, $N_{\text B}$. The whole effect of the
structure of the symmetry group will thus be the simple algebraic prefactor,
$$
\delta_{cd}\tr_{\text F}(Q_a\he Q_b Q_c\he Q_d)=
\frac{N_{\text B}}{N_{\text F}}\delta_{ab}.
$$
The full expression for the inverse boson propagator with one-loop
correction then reads
\begin{multline*}
\D^{-1}(P)=\D_0^{-1}(P)+N\frac{N_{\text B}}{N_{\text F}}
\sumint dK\,dQ\,\D_0(Q)\\
\times\tr\bigl[\gamma_5\G_{11}(K+P)\gamma_5\G_{22}(K)\gamma_5
\G_{11}(K+Q)\gamma_5\G_{22}(K)\bigr],
\end{multline*}
where the subscripts ${}_{11}$ and ${}_{22}$ refer to the matrix structure of
the fermion propagator in the Nambu space. For the sake of brevity we used the
notation for the four-momentum, $P=(i\omega_n,\vec p)$, and the sum-integral,
$$
\sumint dK=T\sum_n\int\frac{d^3\vec k}{(2\pi)^3}.
$$
For zero external momentum the Matsubara sum in the fermion loop may easily be
done and we arrive at the final analytic result,
\begin{widetext}
\begin{align}
\label{rel_propagator}
\D^{-1}(0)&=\D_0^{-1}(0)+N\frac{N_{\text B}}{N_{\text F}}
\sumint dQ\,\D_0(Q)\sum_{e,f=\pm}\int\frac{d^3\vec k}{(2\pi)^3}
\left[1+ef\frac{m^2+\vec k\cdot(\vec k+\vec q)}
{\epsilon_{\vec k}\epsilon_{\vec k+\vec q}}\right]
I(e\xi^e_{\vec k},f\xi^f_{\vec k+\vec q};i\Omega_n),\\
\notag
I(a,b;i\Omega_n)&=\frac1{8a^2}
\frac{\tanh\frac{\beta a}2+\tanh\frac{\beta b}2-\beta
a\cosh^{-2}\frac{\beta a}2}{i\Omega_n+b+a}+
\frac1{8a^2}
\frac{\tanh\frac{\beta a}2-\tanh\frac{\beta b}2}{i\Omega_n+b-a}+
\frac1{4a}\frac{\tanh\frac{\beta a}2+\tanh\frac{\beta b}2}{(i\Omega_n+b+a)^2},
\end{align}
\end{widetext}
with the usual relativistic notation, $\epsilon_{\vec k}=\sqrt{\vec k^2+m^2}$,
$\xi^e_{\vec k}=\epsilon_{\vec k}+e\mu$. Note that upon taking the
nonrelativistic
limit, that is, including only particle degrees of freedom (with $e=f=-$),
shifting the chemical potential to $\mu_{\text{NR}}=\mu-m$, and approximating
the dispersion relation with $\epsilon_{\vec k}=m+\frac{\vec k^2}{2m}$, we
reproduce
the expression for the one-loop propagator correction given by Nikoli\'c and
Sachdev \cite{Nikolic:2007ni}.

\subsection{High-density approximation}
\label{Subsec:HDET}
The nonrelativistic limit of Eq. \eqref{rel_propagator} may be used directly to
calculate the NLO corrections to the critical temperature and chemical potential
via Eq. \eqref{NLO_corr_Tc}. To that end, we need to supplement Eq.
\eqref{rel_propagator} with an analogous one-loop correction to particle
density, which in turn yields the term $\partial_\mu\Omega^{(1)}$ in Eq.
\eqref{NLO_corr_Tc}. This is how the results presented in Fig. \ref{Fig:NR_Tc}
were obtained.

On the contrary, in the full relativistic description the one-loop term in Eq.
\eqref{rel_propagator} is badly divergent. This is, of course, no surprise
since in a relativistic scalar self-interacting field theory the one-loop
tadpole graph has a quadratic divergence. Nevertheless, this divergence has
nothing in common with the many-body physics, and can be removed by
renormalizing the parameters of the theory in the vacuum. In order to avoid
this complication and also the interference of all energy scales from the
high-energy vacuum physics down to the scale of Cooper pairing, we resort to a
high-density approximation \cite{Nardulli:2002ma}, which is appropriate in the
far BCS region where the pairing energy scale is well separated from the Fermi
scale. This approximation is at the MF level known to soften the
ultraviolet divergences and give some cutoff-independent predictions such as the
universal BCS ratio of the gap at zero temperature and the critical temperature
\cite{Bailin:1984bm}.

In our one-loop calculation we have to be more careful. We therefore spell
explicitly all simplifying assumptions that we make. First, we neglect
antiparticle contributions. This can be appropriate even in the
ultrarelativistic limit as long as the pairing gap/critical temperature is much
smaller than the Fermi energy so that the relevant excitations are the
quasiparticles and quasiholes near the Fermi surface. In a strongly coupled
relativistic superconductor, the antiparticle effects may be non-negligible
\cite{Chatterjee:2008dr} but will not change our conclusions qualitatively.

Second, we approximate, as usual, the volume measure for the integral over
fermionic momentum by
$$
\int\frac{d^3\vec k}{(2\pi)^3}\to\mathcal N\int d\xi
\int\frac{d\Omega_{\vec k}}{4\pi},
$$
where $\xi\equiv\xi^-_{\vec k}$ is the energy with respect to the Fermi level
and $\mathcal N=\mu\kf/2\pi^2$ is the density of states on the Fermi surface.
This brings in one subtlety. In a nonrenormalizable theory such as
the NJL model, there is an inherent ambiguity in the way we label internal
propagators of the Feynman graphs with momenta which satisfy momentum
conservation in the interaction vertices. For instance, in the one-loop pairing
susceptibility \eqref{RLOsusc} one often labels the propagators with $\vec
k,\vec k\pm\vec p$ instead of $\vec k+\frac{\vec p}{2},\vec k-\frac{\vec p}{2}$
used here, and imposes a cutoff on the loop momentum $\vec k$. This ambiguity
can be removed by a proper renormalization which makes the graph finite.
However, once we make the above introduced replacement of the integration
measure in the high-density approximation, the new measure is no longer
translationally invariant. That is, \emph{the result depends on the momentum
assignment to the propagators} even in an otherwise finite loop integral. In Eq.
\eqref{RLOsusc} we chose the symmetric momentum assignment because it reduces
cutoff dependence of the result and also leads to a physically intuitive
suppression of the high-momentum pair modes by Pauli blocking.

Third and most importantly, in order to retain in the calculation only the
physically relevant degrees of freedom, we cut off the $\xi$ integration at the
pairing scale, $|\xi|\leq\Lambda$. In quark matter, such a cutoff is effectively
introduced in terms of the momentum-dependent gap function by solving the QCD
Schwinger--Dyson equations \cite{Pisarski:1999tv}. In addition, a cutoff much
smaller than the Fermi energy is necessary in order to make the high-density
approximation consistent \cite{Nardulli:2002ma}. In practice, we verify that
the results are not sensitive to a precise value of the cutoff by doing all
calculations for two different values, $\Lambda=2T_c^{(0)}$ and
$\Lambda=4T_c^{(0)}$.

Fourth, once we restrict ourselves to the low-energy excitations about the
Fermi surface, we expand the dispersion relations in terms of the Fermi
velocity and an ``effective mass''. This allows us to introduce
efficiently dimensionless variables and treat on the same footing the
nonrelativistic (NR) limit of Sec. \ref{Sec:NRgas} as well as the
opposite-extreme, ultrarelativistic (UR) limit of zero fermion mass (which is a
reasonable approximation for quark matter composed solely of the $u$ and $d$
flavors). Concretely, taking the LO critical temperature $T_c^{(0)}$ as a unit
for the energy variables and $T_c^{(0)}/v_{\text F}$, where $v_{\text F}$ is the
Fermi velocity, as a unit for external momentum, the dispersions in the two
opposite limits acquire very similar forms,
\begin{align*}
\bar\xi_{\vec k+\vec p}&=\bar\xi_{\vec k}+
|\bar{\vec p}|\cos\theta+\frac{\bar{\vec p}^2}{4}
\frac{T_c^{(0)}}{\mu},&&\text{NR limit},\\
\bar\xi_{\vec k+\vec p}&=\bar\xi_{\vec k}+
|\bar{\vec p}|\cos\theta+\frac{\bar{\vec p}^2}{2}
\frac{T_c^{(0)}}\mu\sin^2\theta,&&\text{UR limit},
\end{align*}
where $\bar\xi=\xi/T_c^{(0)}$ and $\bar{\vec p}=\vec pv_{\text F}/T_c^{(0)}$,
$\theta$ is the angle between the vectors $\vec k$ and $\vec p$, and the
nonrelativistic chemical potential is understood in the first line, without
specifying the subscript ${}_{\text{NR}}$ in the following. Using this
procedure, all integrals would factorize into a product of powers of the
temperature and Fermi velocity and a universal dimensionless function of the
ratio $T_c^{(0)}/\mu$, were it not for the coupling $g$.

The last step in the construction therefore has to be the renormalization of
the bare coupling. In the context of atomic gases near unitarity it is
customary to do this by fixing the $s$-wave scattering length at zero
momentum in the vacuum as in Eq. \eqref{renorm_prescr}. However, in an
effective description near the Fermi surface, this is no longer convenient. We
therefore renormalize the bare coupling with the help of the gap equation at
zero temperature, which has the same divergence structure as the inverse
propagator \eqref{RLOsusc}. This is effectively done by the replacement
\begin{multline*}
\frac{1-f(\xi_{\vec k_+})-f(\xi_{\vec k_-})}
{i\omega_n-\xi_{\vec k_+}-\xi_{\vec k_-}}\\
\to\frac{1-f(\xi_{\vec k_+})-f(\xi_{\vec k_-})}
{i\omega_n-\xi_{\vec k_+}-\xi_{\vec k_-}}
+\frac{1}{2\sqrt{\xi_{\vec k}^2+\Delta_0^2}}
\end{multline*}
in the inverse propagator \eqref{RLOsusc} in the high-density approximation.
Now everything is expressed in terms of the dimensionless ratios $T_c^{(0)}/\mu$
and $T_c^{(0)}/\Delta_0$. The ratio $T_c^{(0)}/\mu$ is used as the input
parameter which measures the strength of the interaction. The ratio
$T_c^{(0)}/\Delta_0$ is equal to the BCS value
$\frac{e^\gamma}{\pi}\approx0.567$ in the infinite-cutoff limit. With the
explicit cutoff on the $\xi$-integration, we adjust the value of
$\Delta_0$ appropriately in order to ensure that the Goldstone theorem is
satisfied and the propagator \eqref{RLOsusc} has an exactly massless pole.

\begin{figure}
\includegraphics[scale=1]{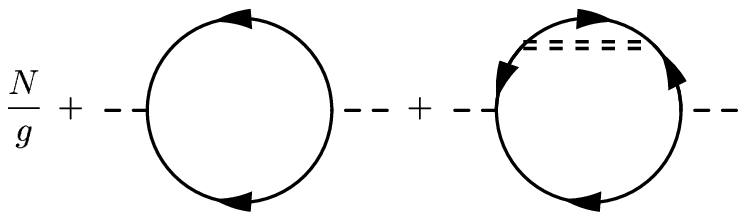}
\caption{Sum of the LO and NLO contributions to the inverse boson propagator.
The asymptotically constant part of the fermion self-energy in the second
diagram may be absorbed in a perturbative renormalization of the Fermi energy.}
\label{Fig:fermi_selfenergy}
\end{figure}

Due to the explicit cutoff, the loop part of the inverse propagator
\eqref{RLOsusc} drops rapidly for external momenta larger than the cutoff as a
result of Pauli blocking. The boson propagator approaches a constant value,
equal to $g/N$. This is natural: At large momentum, pairing fluctuations are
suppressed and the auxiliary field propagator recovers the original contact
four-fermion interaction. In order that we really include just the effect of
the fluctuations of the order parameter, we make in the second term of Eq.
\eqref{rel_propagator} the replacement
$$
\D_0(Q)\to\D_0(Q)-\frac gN.
$$
In terms of Feynman graphs, this means removing from the RPA series of
diagrams the first, constant term, keeping all other terms that involve
multiple rescattering of the two fermions in the pair. Formally, this
subtraction can be justified by observing that the diagram in Fig.
\ref{Fig:box_diagram} may also be viewed as a fermion loop with the insertion
of a one-loop fermion self-energy, see Fig. \ref{Fig:fermi_selfenergy}. The
fermion propagator with the insertion of $g/N$ is nothing else than the
first-order perturbative correction by the contact four-fermion interaction. We
can then pick this constant term out of the one-boson-loop diagram and add it
to the one-fermion-loop MF graph contributing to $\D_0^{-1}(0)$, where it is
absorbed in the perturbative renormalization of the Fermi energy
\footnote{Actually, the first-order perturbative renormalization of the fermion
propagators in the fermion loop in $\D_0^{-1}(0)$ produces two equal terms,
one from each propagator. However, this is exactly what we need because the
loop diagram, being symmetric, carries a symmetry factor $\frac12$.}.

\subsection{Results}
\label{Subsec:rel_results}
\begin{figure}
\includegraphics[scale=1.1]{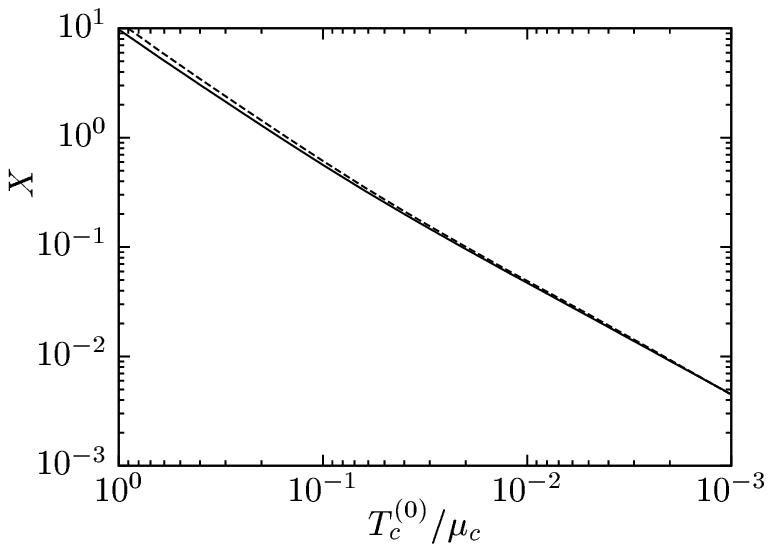}
\caption{Correction to the critical temperature calculated using the
high-density approximation in the \emph{nonrelativistic limit}. The calculation
was done with the cutoff on fermion energy variable set to $2T_c^{(0)}$ (solid
line) and $4T_c^{(0)}$ (dashed line).}
\label{Fig:HDET_NR}
\end{figure}
\begin{figure}
\includegraphics[scale=1.1]{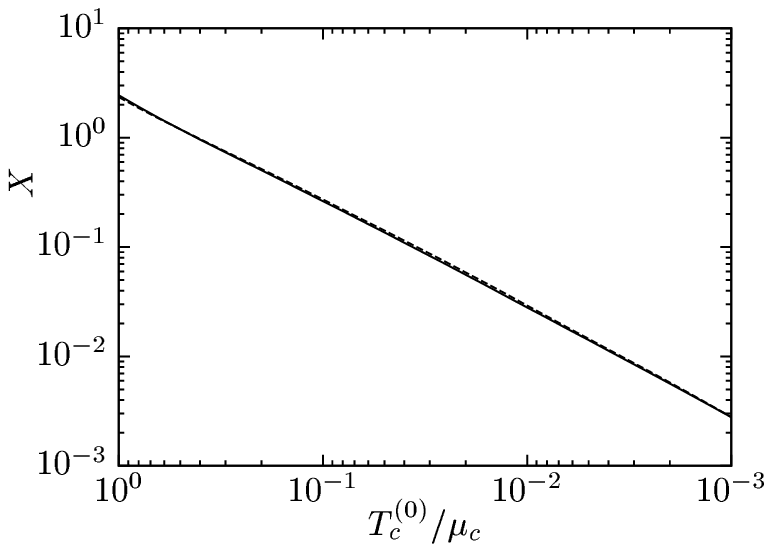}
\caption{Correction to the critical temperature calculated using the
high-density approximation in the \emph{ultrarelativistic limit}. The
calculation was done with the cutoff on fermion energy variable set to
$2T_c^{(0)}$ (solid line) and $4T_c^{(0)}$ (dashed line).}
\label{Fig:HDET_R}
\end{figure}
Recalling finally from the BCS theory that at temperature $T$ different from
the MF critical temperature $T_c^{(0)}$, $\D_0^{-1}(0)$ is equal to $\mathcal
N\log(T/T_c^{(0)})$, setting $\D^{-1}(0)$ to zero leads to the following
compact formula for the critical temperature,
\begin{equation}
T_c=T_c^{(0)}\exp\left[-\frac{N_{\text B}}{N_{\text F}}
X\biggl(\frac{T_c^{(0)}}{\mu}\biggr)\right],
\label{HDET_final_result}
\end{equation}
where $X$ is a dimensionless function (different in the two limits), given by
the integral in Eq. \eqref{rel_propagator} in the dimensionless variables
introduced above.

Eq. \eqref{HDET_final_result} constitutes our final result regarding
relativistic superconductors, written in the most general form: The coupling
constant is completely eliminated in favor of the MF ratio $T_c^{(0)}/\mu$.
Also, the specific form of the pairing channel only enters through the
algebraic factor ${N_{\text B}}/{N_{\text F}}$.

Two comments are in order here. First, Eq. \eqref{HDET_final_result} gives the
correction to the critical temperature at fixed chemical potential; we do not
solve the number equation along with the Thouless criterion to obtain a result
at fixed density. In the context of color-superconducting quark matter, the
(baryon number) chemical potential is usually treated as a free parameter. In
fact, it is a more suitable parameter than the density itself in the case that
the phase diagram involves first-order phase transitions, where the density
becomes discontinuous.

Second, the $1/N$ algorithm based on Eq. \eqref{NLO_corr_Tc} (reduced to a
one-variable problem by fixing the chemical potential) would suggest to
interpret the exponent in Eq. \eqref{HDET_final_result} as the relative
change of the critical temperature, or minus the relative change of the inverse
critical temperature, depending on the choice of variable. In this section, we
used the $1/N$ expansion to derive Eq. \eqref{rel_propagator} as the one-loop
corrected Thouless criterion, and to evaluate the loop correction at the MF
value of the critical temperature. We now go slightly beyond the $1/N$
philosophy in the sense that our result, Eq. \eqref{HDET_final_result}, does
not need any further expansion and defines the critical temperature $T_c$ in an
unambiguous manner. The value of $T_c$ thus calculated is always positive, no
matter how large the loop function $X$ is. (Of course, we would still consider
the used approximation unreliable once $X$ becomes of order one or larger.) In
addition, the temperature correction does not display a finite offset in the
BCS limit, as found in Sec. \ref{Sec:NRgas} and assigned to the $1/N$ expansion
as an artifact. On the contrary, it drops rapidly with decreasing ratio
$T_c^{(0)}/\mu$, as one would naively expect \footnote{As shown by Gorkov and
Melik-Barkhudarov, particle--hole fluctuations lead to a suppression of the gap
and critical temperature even in the BCS limit
\cite{Gorkov:1961gm,Furnstahl:2006pa}. We do not expect this effect to be
reproduced by our calculation since it includes only pair fluctuations
\cite{Diener:2007di}.}.

To complete the discussion of the results, we show in Figs. \ref{Fig:HDET_NR}
and \ref{Fig:HDET_R} the numerically calculated values of the function $X$ in
the NR and UR limits, respectively. For $T_c^{(0)}/\mu$ smaller than about
$0.1$ the functions may be very well approximated by a simple empirical power
law,
\begin{equation}
X_{\text{NR}}\biggl(\frac{T_c^{(0)}}{\mu}\biggr)
\approx5.2\frac{T_c^{(0)}}{\mu},\quad
X_{\text{UR}}\biggl(\frac{T_c^{(0)}}{\mu}\biggr)
\approx2.8\frac{T_c^{(0)}}{\mu},
\label{HDET_limits}
\end{equation}
which can be used for a fast rough estimate of the size of fluctuation effects.

\subsection{Possible impact on QCD phase diagram}
\begin{figure*}
\begin{center}
\includegraphics[scale=0.2]{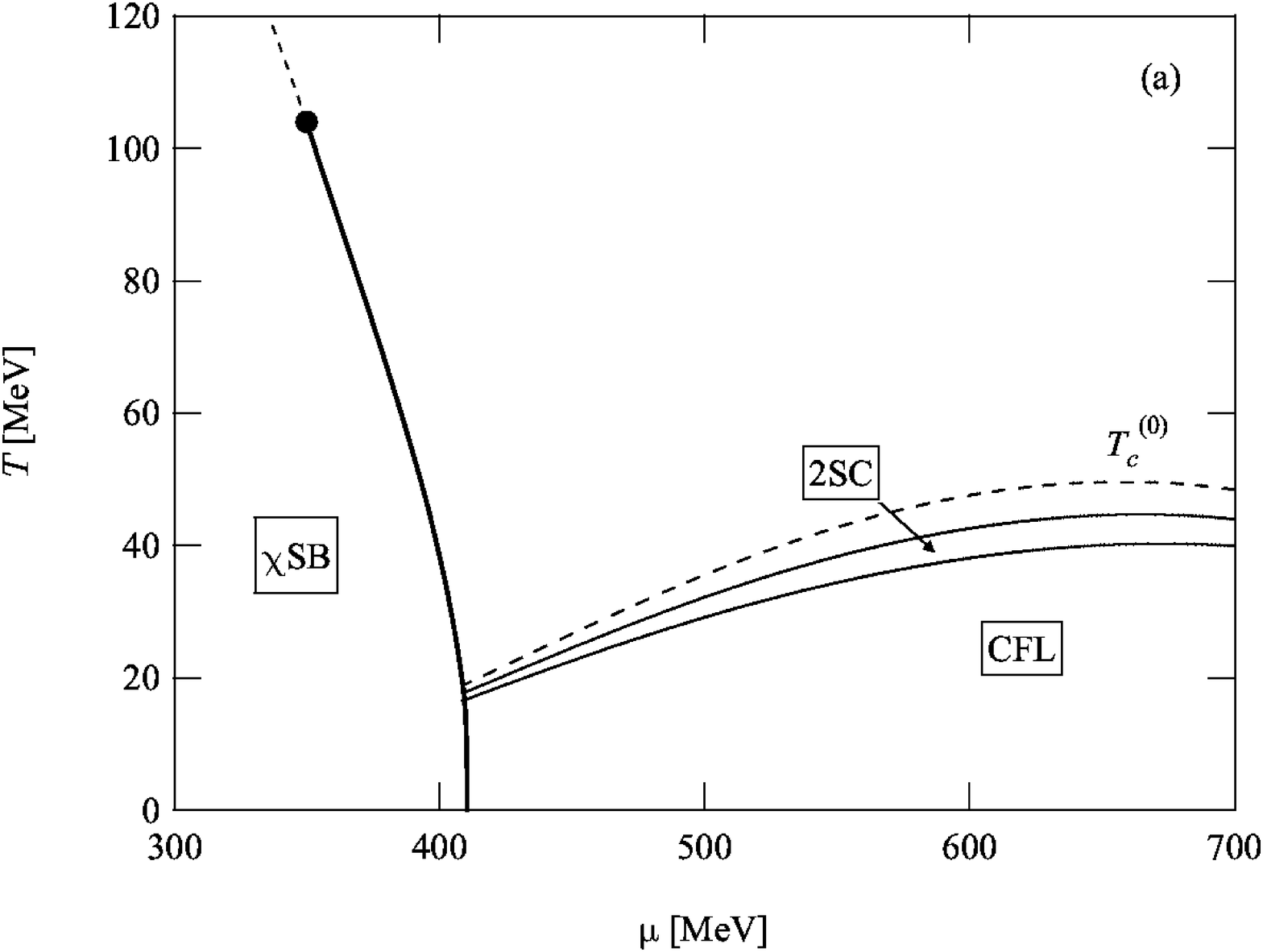}\hskip2em
\includegraphics[scale=0.2]{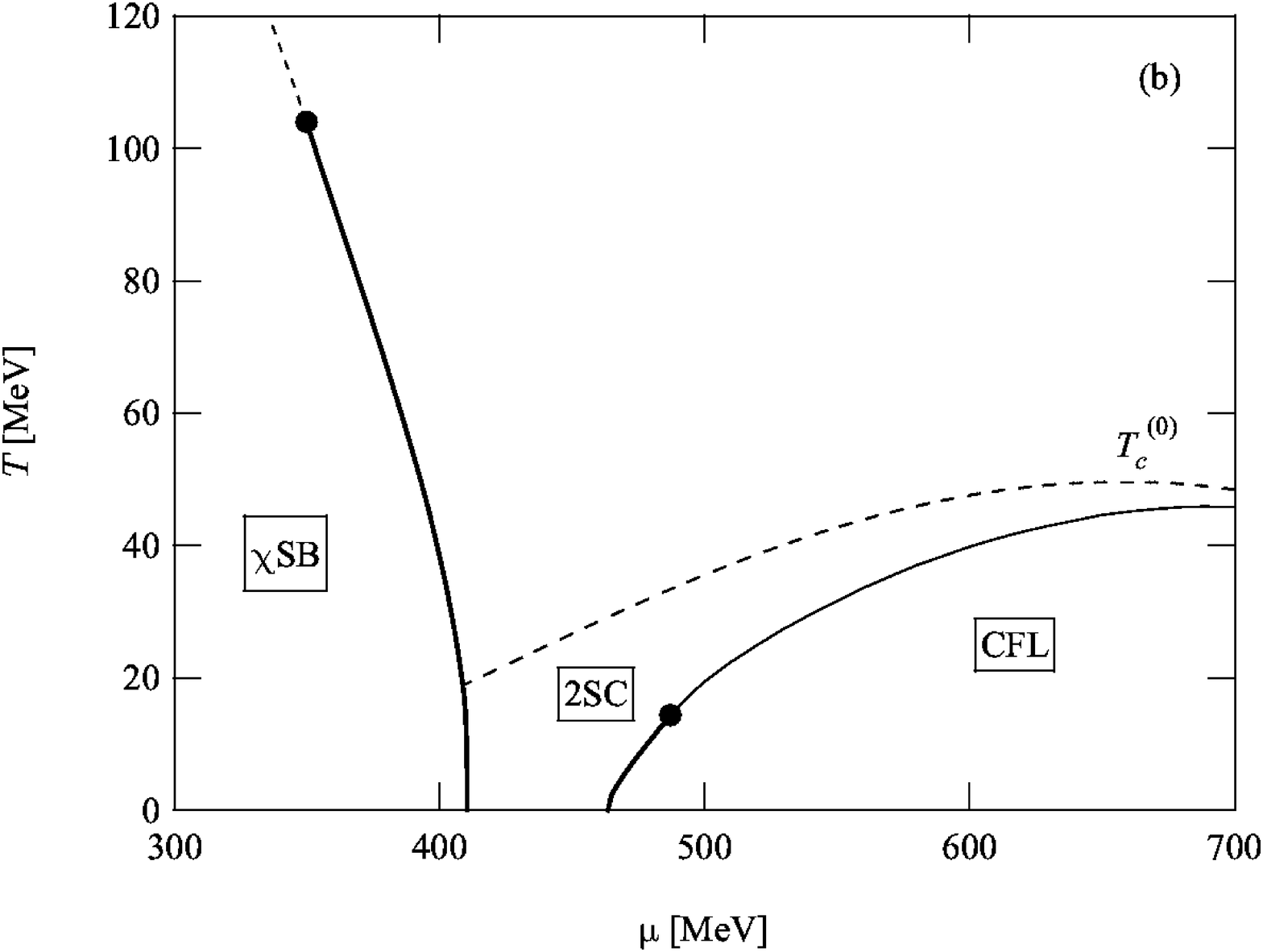}
\end{center}
\caption{(a) Phase diagram at vanishing strange quark mass $M_s=0$ but
\emph{with} quantum fluctuation effects. (b) The same phase diagram for
$M_s=200$~MeV \emph{without} fluctuation effects. $T_{c}^{(0)}$ is the
critical temperature at the leading order in $1/N$.}
\label{Fig:phase_diagram}
\end{figure*}
Finally, we wish to illustrate a possible impact of fluctuations on the QCD
phase diagram. In Fig. \ref{Fig:phase_diagram} we display the phase diagrams
from a simple NJL-model calculation; the model is the same one as adopted in
Ref. \cite{Abuki:2005ms} with the diquark coupling $G_d$ chosen such that
the CFL gap is $50\,\text{MeV}$ at $\mu=500\,\text{MeV}$ in the chiral
$\mathrm{SU}(3)$ limit. Fig. \ref{Fig:phase_diagram} (a) shows the phase diagram
at vanishing strange quark mass, i.e., in the chiral $\mathrm{SU}(3)$ limit.
The dashed line is the critical temperature at the leading order in $1/N$
expansion, and the suppression of the critical temperature due to fluctuation
effects at NLO is included by means of the analytic formula
\eqref{HDET_final_result}. Interestingly, because $N_{\text B}/N_{\text F}=1$
in the CFL case while it is just $1/2$ in the 2SC phase, we expect a finite
region with 2SC pairing below the normal phase even in the chiral
$\mathrm{SU}(3)$ limit.

However, since the approximation which led to Eq. \eqref{HDET_final_result} is
only valid in the high-density regime where $T_{c}^{(0)}/\mu$ is small, we have
to keep in mind that the estimates are not quantitative at low density. It is
also important to note that we derived the shift of critical temperature of the
CFL phase, taking into account the fluctuations in the normal phase. Now that
we know that the 2SC phase interposes between the normal and CFL phases, it
would be more appropriate to somehow take into account the fluctuations within
the 2SC phase for a more realistic estimate of the temperature of the phase
transition between the CFL and 2SC phases.

Fig. \ref{Fig:phase_diagram} (b) shows the phase diagram at a finite
strange quark mass, but without quantum fluctuation effects. The strange quark
mass is set to $M_s=200\,\text{MeV}$, and is for simplicity treated as an
external parameter rather than a dynamical one. Also, the intricate charge
neutrality effects are ignored here because they are not important for our
purposes; we just remark that the charge neutrality constraints bring
fine splittings to the $T_c$'s resulting in the appearance of tiny regions with
dSC or uSC pairing \cite{Abuki:2005ms,Iida:2003cc}. Comparing the two plots, we
can see that the quantum fluctuation effects may play a significant role at high
density which may be similar to that of the strange quark mass. This can be
understood in a model-independent way as follows. Within a weak-coupling
Ginzburg--Landau approach, it was shown that the strange quark mass and charge
neutrality result in shifts in the melting temperatures $T_i$ of the order
parameters $\Delta_i$ of the CFL phase of the order \cite{Iida:2003cc}
$$
\frac{\delta T_i}{T_c^{(0)}}\sim-\frac{M_s^2}{8\mu^2}
\log\left(\frac{\mu}{T_c^{(0)}}\right).
$$
These corrections die rapidly as $(M_s/\mu)^2$ at large $\mu$, while the
correction to $T_c^{(0)}$ from order parameter fluctuations prevails as
it behaves asymptotically like $T_c^{(0)}/\mu$. Considering in addition the
fact that $M_s$ is a decreasing function of $\mu$ whereas the superconducting
gap turns out to increase as $\mu\to\infty$ \cite{Son:1998uk,Schaefer:1999jg},
we conclude that the quantum fluctuation is more important than the effect of a
strange quark mass at high density.

In a more realistic situation, there would be another source of fluctuations
from thermal configurations of gauge fields \cite{Giannakis:2004xt} which makes
the superconducting--normal transition first order. The shift of the critical
temperature turns out to be positive and is estimated to be proportional to
the QCD coupling $g$ in the weak-coupling regime \cite{Giannakis:2004xt}.
Therefore, in realistic quark matter the order parameter and gauge field
fluctuations will compete each other. This issue certainly deserves a further
study in future.

\section{Summary and conclusions}
\label{Sec:conclusions}
We have investigated the $1/N$ expansion for strongly interacting Fermi
systems, proposed recently \cite{Nikolic:2007ni,Veillette:2007ve}. We first
studied in detail the case of nonrelativistic Fermi gas near unitarity,
extending the previous results by the calculation of the critical temperature
off the unitarity. Even though the $1/N$ expansion can give a result for the
critical temperature at unitarity which is very close to the prediction from
Monte Carlo simulations \cite{Burovski:2006bu,Burovski:2008bu}, there is an
inherent ambiguity due to the choice of observable to generate (and truncate)
the $1/N$ series. This ambiguity makes the $1/N$ expansion in the
current setting useless when the corrections to the MF theory are large, in
particular on the BEC side of the crossover.

We paid particular attention to the evolution of the fluctuation corrections in
the BCS regime, where they are expected to be small, and the $1/N$ series thus
to converge fast. We showed that the next-to-leading order in the $1/N$
expansion reproduces the well-known perturbative correction to the chemical
potential up to second order. As far as the critical temperature is concerned,
the fluctuation correction indeed decreases at weak coupling, but leaves a
finite offset in the $\kf a\to0-$ limit. We argued that this is likely to be an
artifact of the $1/N$ expansion.

In Sec. \ref{Sec:rel_matter} we applied the idea to strongly coupled
relativistic superconductors, having in mind in particular
color-superconducting dense quark matter. We used a simple class of NJL-type
models and resorted to a high-density approximation in order to avoid
conceptual difficulties associated with renormalization and entanglement of
several energy scales. Our results are summarized in Eqs.
\eqref{HDET_final_result} and \eqref{HDET_limits} and Figures \ref{Fig:HDET_NR}
and \ref{Fig:HDET_R}. They are physically intuitive in the sense that the
fluctuation corrections are small in the BCS limit, decreasing linearly with
the ratio $T_c^{(0)}/\mu$, i.e., exponentially with the inverse coupling,
and become large as $T_c^{(0)}/\mu$ approaches the order of $0.1$ and further
grows.

In particular for typical color superconductors, the corrections to
critical temperature are expected to be as large as tens percent. Another
important conclusion is that, within the simple setting used here, the
fluctuation corrections are expressed as a universal function of the
dimensionless ratio $T_c^{(0)}/\mu$.  The whole dependence on the symmetry
structure of the pairing is encoded in an algebraic prefactor, which counts
the number of bosonic and fermionic degrees of freedom. This makes it
straightforward to compare the effects for different competing superconducting
phases, and thus estimate the impact of the order parameter fluctuations on the
phase diagram such as in Fig. \ref{Fig:phase_diagram}. Upon this investigation
of the fluctuation corrections within a simple model, we plan, in our future
work, to include the effects of chemical potential mismatch and color neutrality
in order to obtain a more realistic description of quark matter.

Finally, we would like to stress the conceptual simplicity of this approach to
order parameter fluctuations. We do not need to solve a complicated set of
self-consistent integral equations like in other techniques going beyond the MF
approximation, such as the Cornwall--Jackiw--Tomboulis one. Instead, one just
has to evaluate a single multidimensional sum-integral. We therefore believe
that the $1/N$ expansion may provide an efficient tool to determine the
fluctuation effects in such strongly-coupled systems as, for instance, the color
superconductors.

\begin{acknowledgments}
The authors are grateful to J. Ho\v{s}ek and D. H. Rischke for fruitful
discussions. The present work was supported by the Alexander
von Humboldt Foundation. Numerical calculations were performed using the
facilities of the Frankfurt Center for Scientific Computing, and in part on
Altix3700 at the Yukawa Institute for Theoretical Physics of Kyoto University.
\end{acknowledgments}


\begin{thebibliography}{58}
\expandafter\ifx\csname natexlab\endcsname\relax\def\natexlab#1{#1}\fi
\expandafter\ifx\csname bibnamefont\endcsname\relax
  \def\bibnamefont#1{#1}\fi
\expandafter\ifx\csname bibfnamefont\endcsname\relax
  \def\bibfnamefont#1{#1}\fi
\expandafter\ifx\csname citenamefont\endcsname\relax
  \def\citenamefont#1{#1}\fi
\expandafter\ifx\csname url\endcsname\relax
  \def\url#1{\texttt{#1}}\fi
\expandafter\ifx\csname urlprefix\endcsname\relax\def\urlprefix{URL }\fi
\providecommand{\bibinfo}[2]{#2}
\providecommand{\eprint}[2][]{\url{#2}}

\bibitem[{\citenamefont{Eagles}(1969)}]{Eagles:1969ea}
\bibinfo{author}{\bibfnamefont{D.~M.} \bibnamefont{Eagles}},
  \bibinfo{journal}{Phys. Rev.} \textbf{\bibinfo{volume}{186}},
  \bibinfo{pages}{456} (\bibinfo{year}{1969}).

\bibitem[{\citenamefont{Leggett}(1980)}]{Leggett:1980le}
\bibinfo{author}{\bibfnamefont{A.~J.} \bibnamefont{Leggett}},
  \bibinfo{journal}{J. Phys. Colloques} \textbf{\bibinfo{volume}{41}},
  \bibinfo{pages}{19} (\bibinfo{year}{1980}).

\bibitem[{\citenamefont{{Nozi\`eres} and Schmitt-Rink}(1985)}]{Nozieres:1985no}
\bibinfo{author}{\bibfnamefont{P.}~\bibnamefont{{Nozi\`eres}}}
  \bibnamefont{and}
  \bibinfo{author}{\bibfnamefont{S.}~\bibnamefont{Schmitt-Rink}},
  \bibinfo{journal}{J. Low Temp. Phys.} \textbf{\bibinfo{volume}{59}},
  \bibinfo{pages}{195} (\bibinfo{year}{1985}).

\bibitem[{\citenamefont{Regal et~al.}(2004)\citenamefont{Regal, Greiner, and
  Jin}}]{Regal:2004re}
\bibinfo{author}{\bibfnamefont{C.~A.} \bibnamefont{Regal}},
  \bibinfo{author}{\bibfnamefont{M.}~\bibnamefont{Greiner}}, \bibnamefont{and}
  \bibinfo{author}{\bibfnamefont{D.~S.} \bibnamefont{Jin}},
  \bibinfo{journal}{Phys. Rev. Lett.} \textbf{\bibinfo{volume}{92}},
  \bibinfo{pages}{040403} (\bibinfo{year}{2004}), \eprint{cond-mat/0401554}.

\bibitem[{\citenamefont{Bartenstein et~al.}(2004)\citenamefont{Bartenstein,
  Altmeyer, Riedl, Jochim, Chin, Hecker~Denschlag, and
  Grimm}}]{Bartenstein:2004ba}
\bibinfo{author}{\bibfnamefont{M.}~\bibnamefont{Bartenstein}},
  \bibinfo{author}{\bibfnamefont{A.}~\bibnamefont{Altmeyer}},
  \bibinfo{author}{\bibfnamefont{S.}~\bibnamefont{Riedl}},
  \bibinfo{author}{\bibfnamefont{S.}~\bibnamefont{Jochim}},
  \bibinfo{author}{\bibfnamefont{C.}~\bibnamefont{Chin}},
  \bibinfo{author}{\bibfnamefont{J.}~\bibnamefont{Hecker~Denschlag}},
  \bibnamefont{and} \bibinfo{author}{\bibfnamefont{R.}~\bibnamefont{Grimm}},
  \bibinfo{journal}{Phys. Rev. Lett.} \textbf{\bibinfo{volume}{92}},
  \bibinfo{pages}{120401} (\bibinfo{year}{2004}), \eprint{cond-mat/0401109}.

\bibitem[{\citenamefont{Zwierlein et~al.}(2004)\citenamefont{Zwierlein, Stan,
  Schunck, Raupach, Kerman, and Ketterle}}]{Zwierlein:2004zw}
\bibinfo{author}{\bibfnamefont{M.~W.} \bibnamefont{Zwierlein}},
  \bibinfo{author}{\bibfnamefont{C.~A.} \bibnamefont{Stan}},
  \bibinfo{author}{\bibfnamefont{C.~H.} \bibnamefont{Schunck}},
  \bibinfo{author}{\bibfnamefont{S.~M.~F.} \bibnamefont{Raupach}},
  \bibinfo{author}{\bibfnamefont{A.~J.} \bibnamefont{Kerman}},
  \bibnamefont{and} \bibinfo{author}{\bibfnamefont{W.}~\bibnamefont{Ketterle}},
  \bibinfo{journal}{Phys. Rev. Lett.} \textbf{\bibinfo{volume}{92}},
  \bibinfo{pages}{120403} (\bibinfo{year}{2004}), \eprint{cond-mat/0403049}.

\bibitem[{\citenamefont{Bourdel et~al.}(2004)\citenamefont{Bourdel, Khaykovich,
  Cubizolles, Zhang, Chevy, Teichmann, Tarruell, Kokkelmans, and
  Salomon}}]{Bourdel:2004bo}
\bibinfo{author}{\bibfnamefont{T.}~\bibnamefont{Bourdel}},
  \bibinfo{author}{\bibfnamefont{L.}~\bibnamefont{Khaykovich}},
  \bibinfo{author}{\bibfnamefont{J.}~\bibnamefont{Cubizolles}},
  \bibinfo{author}{\bibfnamefont{J.}~\bibnamefont{Zhang}},
  \bibinfo{author}{\bibfnamefont{F.}~\bibnamefont{Chevy}},
  \bibinfo{author}{\bibfnamefont{M.}~\bibnamefont{Teichmann}},
  \bibinfo{author}{\bibfnamefont{L.}~\bibnamefont{Tarruell}},
  \bibinfo{author}{\bibfnamefont{S.~J. J. M.~F.} \bibnamefont{Kokkelmans}},
  \bibnamefont{and} \bibinfo{author}{\bibfnamefont{C.}~\bibnamefont{Salomon}},
  \bibinfo{journal}{Phys. Rev. Lett.} \textbf{\bibinfo{volume}{93}},
  \bibinfo{pages}{050401} (\bibinfo{year}{2004}), \eprint{cond-mat/0403091}.

\bibitem[{\citenamefont{Son}(2008)}]{Son:2008ye}
\bibinfo{author}{\bibfnamefont{D.~T.} \bibnamefont{Son}},
  \bibinfo{journal}{Phys. Rev.} \textbf{\bibinfo{volume}{D78}},
  \bibinfo{pages}{046003} (\bibinfo{year}{2008}), \eprint{arXiv:0804.3972
  [hep-th]}.

\bibitem[{\citenamefont{Balasubramanian and
  McGreevy}(2008)}]{Balasubramanian:2008dm}
\bibinfo{author}{\bibfnamefont{K.}~\bibnamefont{Balasubramanian}}
  \bibnamefont{and} \bibinfo{author}{\bibfnamefont{J.}~\bibnamefont{McGreevy}},
  \bibinfo{journal}{Phys. Rev. Lett.} \textbf{\bibinfo{volume}{101}},
  \bibinfo{pages}{061601} (\bibinfo{year}{2008}), \eprint{arXiv:0804.4053
  [hep-th]}.

\bibitem[{\citenamefont{Haussmann}(1994)}]{Haussmann:1994ha}
\bibinfo{author}{\bibfnamefont{R.}~\bibnamefont{Haussmann}},
  \bibinfo{journal}{Phys. Rev.} \textbf{\bibinfo{volume}{B49}},
  \bibinfo{pages}{12975} (\bibinfo{year}{1994}).

\bibitem[{\citenamefont{Haussmann et~al.}(2007)\citenamefont{Haussmann,
  Rantner, Cerrito, and Zwerger}}]{Haussmann:2007ha}
\bibinfo{author}{\bibfnamefont{R.}~\bibnamefont{Haussmann}},
  \bibinfo{author}{\bibfnamefont{W.}~\bibnamefont{Rantner}},
  \bibinfo{author}{\bibfnamefont{S.}~\bibnamefont{Cerrito}}, \bibnamefont{and}
  \bibinfo{author}{\bibfnamefont{W.}~\bibnamefont{Zwerger}},
  \bibinfo{journal}{Phys. Rev.} \textbf{\bibinfo{volume}{A75}},
  \bibinfo{pages}{023610} (\bibinfo{year}{2007}), \eprint{cond-mat/0608282}.

\bibitem[{\citenamefont{Heiselberg}(2001)}]{Heiselberg:2001he}
\bibinfo{author}{\bibfnamefont{H.}~\bibnamefont{Heiselberg}},
  \bibinfo{journal}{Phys. Rev.} \textbf{\bibinfo{volume}{A63}},
  \bibinfo{pages}{043606} (\bibinfo{year}{2001}), \eprint{cond-mat/0002056}.

\bibitem[{\citenamefont{Nussinov and Nussinov}(2006)}]{Nussinov:2006nu}
\bibinfo{author}{\bibfnamefont{Z.}~\bibnamefont{Nussinov}} \bibnamefont{and}
  \bibinfo{author}{\bibfnamefont{S.}~\bibnamefont{Nussinov}},
  \bibinfo{journal}{Phys. Rev.} \textbf{\bibinfo{volume}{A74}},
  \bibinfo{pages}{053622} (\bibinfo{year}{2006}), \eprint{cond-mat/0410597}.

\bibitem[{\citenamefont{Nishida and Son}(2006)}]{Nishida:2006br}
\bibinfo{author}{\bibfnamefont{Y.}~\bibnamefont{Nishida}} \bibnamefont{and}
  \bibinfo{author}{\bibfnamefont{D.~T.} \bibnamefont{Son}},
  \bibinfo{journal}{Phys. Rev. Lett.} \textbf{\bibinfo{volume}{97}},
  \bibinfo{pages}{050403} (\bibinfo{year}{2006}), \eprint{cond-mat/0604500}.

\bibitem[{\citenamefont{Chen and Nakano}(2007)}]{Chen:2006wx}
\bibinfo{author}{\bibfnamefont{J.-W.} \bibnamefont{Chen}} \bibnamefont{and}
  \bibinfo{author}{\bibfnamefont{E.}~\bibnamefont{Nakano}},
  \bibinfo{journal}{Phys. Rev.} \textbf{\bibinfo{volume}{A75}},
  \bibinfo{pages}{043620} (\bibinfo{year}{2007}), \eprint{cond-mat/0610011}.

\bibitem[{\citenamefont{{Nikoli\'c} and Sachdev}(2007)}]{Nikolic:2007ni}
\bibinfo{author}{\bibfnamefont{P.}~\bibnamefont{{Nikoli\'c}}} \bibnamefont{and}
  \bibinfo{author}{\bibfnamefont{S.}~\bibnamefont{Sachdev}},
  \bibinfo{journal}{Phys. Rev.} \textbf{\bibinfo{volume}{A75}},
  \bibinfo{pages}{033608} (\bibinfo{year}{2007}).

\bibitem[{\citenamefont{Veillette et~al.}(2007)\citenamefont{Veillette, Sheehy,
  and Radzihovsky}}]{Veillette:2007ve}
\bibinfo{author}{\bibfnamefont{M.~Y.} \bibnamefont{Veillette}},
  \bibinfo{author}{\bibfnamefont{D.~E.} \bibnamefont{Sheehy}},
  \bibnamefont{and}
  \bibinfo{author}{\bibfnamefont{L.}~\bibnamefont{Radzihovsky}},
  \bibinfo{journal}{Phys. Rev.} \textbf{\bibinfo{volume}{A75}},
  \bibinfo{pages}{043614} (\bibinfo{year}{2007}).

\bibitem[{\citenamefont{Chen et~al.}(2005)\citenamefont{Chen, Stajic, Tan, and
  Levin}}]{Chen:2005ch}
\bibinfo{author}{\bibfnamefont{Q.}~\bibnamefont{Chen}},
  \bibinfo{author}{\bibfnamefont{J.}~\bibnamefont{Stajic}},
  \bibinfo{author}{\bibfnamefont{S.}~\bibnamefont{Tan}}, \bibnamefont{and}
  \bibinfo{author}{\bibfnamefont{K.}~\bibnamefont{Levin}},
  \bibinfo{journal}{Phys. Rept.} \textbf{\bibinfo{volume}{412}},
  \bibinfo{pages}{1} (\bibinfo{year}{2005}).

\bibitem[{\citenamefont{Carlson et~al.}(2003)\citenamefont{Carlson, Chang,
  Pandharipande, and Schmidt}}]{Carlson:2003ca}
\bibinfo{author}{\bibfnamefont{J.}~\bibnamefont{Carlson}},
  \bibinfo{author}{\bibfnamefont{S.-Y.} \bibnamefont{Chang}},
  \bibinfo{author}{\bibfnamefont{V.~R.} \bibnamefont{Pandharipande}},
  \bibnamefont{and} \bibinfo{author}{\bibfnamefont{K.~E.}
  \bibnamefont{Schmidt}}, \bibinfo{journal}{Phys. Rev. Lett.}
  \textbf{\bibinfo{volume}{91}}, \bibinfo{pages}{050401}
  (\bibinfo{year}{2003}), \eprint{physics/0303094}.

\bibitem[{\citenamefont{Bulgac et~al.}(2006)\citenamefont{Bulgac, Drut, and
  Magierski}}]{Bulgac:2005pj}
\bibinfo{author}{\bibfnamefont{A.}~\bibnamefont{Bulgac}},
  \bibinfo{author}{\bibfnamefont{J.~E.} \bibnamefont{Drut}}, \bibnamefont{and}
  \bibinfo{author}{\bibfnamefont{P.}~\bibnamefont{Magierski}},
  \bibinfo{journal}{Phys. Rev. Lett.} \textbf{\bibinfo{volume}{96}},
  \bibinfo{pages}{090404} (\bibinfo{year}{2006}), \eprint{cond-mat/0505374}.

\bibitem[{\citenamefont{Burovski et~al.}(2006)\citenamefont{Burovski,
  Prokof'ev, Svistunov, and Troyer}}]{Burovski:2006bu}
\bibinfo{author}{\bibfnamefont{E.}~\bibnamefont{Burovski}},
  \bibinfo{author}{\bibfnamefont{N.}~\bibnamefont{Prokof'ev}},
  \bibinfo{author}{\bibfnamefont{B.}~\bibnamefont{Svistunov}},
  \bibnamefont{and} \bibinfo{author}{\bibfnamefont{M.}~\bibnamefont{Troyer}},
  \bibinfo{journal}{Phys. Rev. Lett.} \textbf{\bibinfo{volume}{96}},
  \bibinfo{pages}{160402} (\bibinfo{year}{2006}), \eprint{cond-mat/0602224}.

\bibitem[{\citenamefont{Burovski et~al.}(2008)\citenamefont{Burovski, Kozik,
  Prokof'ev, Svistunov, and Troyer}}]{Burovski:2008bu}
\bibinfo{author}{\bibfnamefont{E.}~\bibnamefont{Burovski}},
  \bibinfo{author}{\bibfnamefont{E.}~\bibnamefont{Kozik}},
  \bibinfo{author}{\bibfnamefont{N.}~\bibnamefont{Prokof'ev}},
  \bibinfo{author}{\bibfnamefont{B.}~\bibnamefont{Svistunov}},
  \bibnamefont{and} \bibinfo{author}{\bibfnamefont{M.}~\bibnamefont{Troyer}},
  \bibinfo{journal}{Phys. Rev. Lett.} \textbf{\bibinfo{volume}{101}},
  \bibinfo{pages}{090402} (\bibinfo{year}{2008}), \eprint{arXiv:0805.3047
  [cond-mat.str-el]}.

\bibitem[{\citenamefont{Nambu and
  Jona-Lasinio}(1961{\natexlab{a}})}]{Nambu:1961tp}
\bibinfo{author}{\bibfnamefont{Y.}~\bibnamefont{Nambu}} \bibnamefont{and}
  \bibinfo{author}{\bibfnamefont{G.}~\bibnamefont{Jona-Lasinio}},
  \bibinfo{journal}{Phys. Rev.} \textbf{\bibinfo{volume}{122}},
  \bibinfo{pages}{345} (\bibinfo{year}{1961}{\natexlab{a}}).

\bibitem[{\citenamefont{Nambu and
  Jona-Lasinio}(1961{\natexlab{b}})}]{Nambu:1961fr}
\bibinfo{author}{\bibfnamefont{Y.}~\bibnamefont{Nambu}} \bibnamefont{and}
  \bibinfo{author}{\bibfnamefont{G.}~\bibnamefont{Jona-Lasinio}},
  \bibinfo{journal}{Phys. Rev.} \textbf{\bibinfo{volume}{124}},
  \bibinfo{pages}{246} (\bibinfo{year}{1961}{\natexlab{b}}).

\bibitem[{\citenamefont{Barrois}(1977)}]{Barrois:1977xd}
\bibinfo{author}{\bibfnamefont{B.~C.} \bibnamefont{Barrois}},
  \bibinfo{journal}{Nucl. Phys.} \textbf{\bibinfo{volume}{B129}},
  \bibinfo{pages}{390} (\bibinfo{year}{1977}).

\bibitem[{\citenamefont{Frautschi}(1980)}]{Frautschi:1978rz}
\bibinfo{author}{\bibfnamefont{S.~C.} \bibnamefont{Frautschi}}
  (\bibinfo{year}{1980}), \bibinfo{note}{in Workshop on Hadronic Matter at
  Extreme Energy Density, Erice, Italy, Oct 13-21, 1978}.

\bibitem[{\citenamefont{Alford et~al.}(2007)\citenamefont{Alford, Schmitt,
  Rajagopal, and Schaefer}}]{Alford:2007rw}
\bibinfo{author}{\bibfnamefont{M.~G.} \bibnamefont{Alford}},
  \bibinfo{author}{\bibfnamefont{A.}~\bibnamefont{Schmitt}},
  \bibinfo{author}{\bibfnamefont{K.}~\bibnamefont{Rajagopal}},
  \bibnamefont{and} \bibinfo{author}{\bibfnamefont{T.}~\bibnamefont{Schaefer}}
  (\bibinfo{year}{2007}), \eprint{arXiv:0709.4635 [hep-ph]}.

\bibitem[{\citenamefont{Babaev}(2001)}]{Babaev:1999iz}
\bibinfo{author}{\bibfnamefont{E.}~\bibnamefont{Babaev}},
  \bibinfo{journal}{Int. J. Mod. Phys.} \textbf{\bibinfo{volume}{A16}},
  \bibinfo{pages}{1175} (\bibinfo{year}{2001}), \eprint{hep-th/9909052}.

\bibitem[{\citenamefont{Matsuzaki}(2000)}]{Matsuzaki:1999ww}
\bibinfo{author}{\bibfnamefont{M.}~\bibnamefont{Matsuzaki}},
  \bibinfo{journal}{Phys. Rev.} \textbf{\bibinfo{volume}{D62}},
  \bibinfo{pages}{017501} (\bibinfo{year}{2000}), \eprint{hep-ph/9910541}.

\bibitem[{\citenamefont{Abuki et~al.}(2002)\citenamefont{Abuki, Hatsuda, and
  Itakura}}]{Abuki:2001be}
\bibinfo{author}{\bibfnamefont{H.}~\bibnamefont{Abuki}},
  \bibinfo{author}{\bibfnamefont{T.}~\bibnamefont{Hatsuda}}, \bibnamefont{and}
  \bibinfo{author}{\bibfnamefont{K.}~\bibnamefont{Itakura}},
  \bibinfo{journal}{Phys. Rev.} \textbf{\bibinfo{volume}{D65}},
  \bibinfo{pages}{074014} (\bibinfo{year}{2002}), \eprint{hep-ph/0109013}.

\bibitem[{\citenamefont{Kitazawa et~al.}(2004)\citenamefont{Kitazawa, Koide,
  Kunihiro, and Nemoto}}]{Kitazawa:2003cs}
\bibinfo{author}{\bibfnamefont{M.}~\bibnamefont{Kitazawa}},
  \bibinfo{author}{\bibfnamefont{T.}~\bibnamefont{Koide}},
  \bibinfo{author}{\bibfnamefont{T.}~\bibnamefont{Kunihiro}}, \bibnamefont{and}
  \bibinfo{author}{\bibfnamefont{Y.}~\bibnamefont{Nemoto}},
  \bibinfo{journal}{Phys. Rev.} \textbf{\bibinfo{volume}{D70}},
  \bibinfo{pages}{056003} (\bibinfo{year}{2004}), \eprint{hep-ph/0309026}.

\bibitem[{\citenamefont{Nawa et~al.}(2006)\citenamefont{Nawa, Nakano, and
  Yabu}}]{Nawa:2005sb}
\bibinfo{author}{\bibfnamefont{K.}~\bibnamefont{Nawa}},
  \bibinfo{author}{\bibfnamefont{E.}~\bibnamefont{Nakano}}, \bibnamefont{and}
  \bibinfo{author}{\bibfnamefont{H.}~\bibnamefont{Yabu}},
  \bibinfo{journal}{Phys. Rev.} \textbf{\bibinfo{volume}{D74}},
  \bibinfo{pages}{034017} (\bibinfo{year}{2006}), \eprint{hep-ph/0509029}.

\bibitem[{\citenamefont{Kitazawa et~al.}(2008)\citenamefont{Kitazawa, Rischke,
  and Shovkovy}}]{Kitazawa:2007zs}
\bibinfo{author}{\bibfnamefont{M.}~\bibnamefont{Kitazawa}},
  \bibinfo{author}{\bibfnamefont{D.~H.} \bibnamefont{Rischke}},
  \bibnamefont{and} \bibinfo{author}{\bibfnamefont{I.~A.}
  \bibnamefont{Shovkovy}}, \bibinfo{journal}{Phys. Lett.}
  \textbf{\bibinfo{volume}{B663}}, \bibinfo{pages}{228} (\bibinfo{year}{2008}),
  \eprint{arXiv:0709.2235 [hep-ph]}.

\bibitem[{\citenamefont{Nishida and Abuki}(2005)}]{Nishida:2005ds}
\bibinfo{author}{\bibfnamefont{Y.}~\bibnamefont{Nishida}} \bibnamefont{and}
  \bibinfo{author}{\bibfnamefont{H.}~\bibnamefont{Abuki}},
  \bibinfo{journal}{Phys. Rev.} \textbf{\bibinfo{volume}{D72}},
  \bibinfo{pages}{096004} (\bibinfo{year}{2005}), \eprint{hep-ph/0504083}.

\bibitem[{\citenamefont{Abuki}(2007)}]{Abuki:2006dv}
\bibinfo{author}{\bibfnamefont{H.}~\bibnamefont{Abuki}},
  \bibinfo{journal}{Nucl. Phys.} \textbf{\bibinfo{volume}{A791}},
  \bibinfo{pages}{117} (\bibinfo{year}{2007}), \eprint{hep-ph/0605081}.

\bibitem[{\citenamefont{Deng et~al.}(2007)\citenamefont{Deng, Schmitt, and
  Wang}}]{Deng:2006ed}
\bibinfo{author}{\bibfnamefont{J.}~\bibnamefont{Deng}},
  \bibinfo{author}{\bibfnamefont{A.}~\bibnamefont{Schmitt}}, \bibnamefont{and}
  \bibinfo{author}{\bibfnamefont{Q.}~\bibnamefont{Wang}},
  \bibinfo{journal}{Phys. Rev.} \textbf{\bibinfo{volume}{D76}},
  \bibinfo{pages}{034013} (\bibinfo{year}{2007}), \eprint{nucl-th/0611097}.

\bibitem[{\citenamefont{He and Zhuang}(2007{\natexlab{a}})}]{He:2007kd}
\bibinfo{author}{\bibfnamefont{L.}~\bibnamefont{He}} \bibnamefont{and}
  \bibinfo{author}{\bibfnamefont{P.}~\bibnamefont{Zhuang}},
  \bibinfo{journal}{Phys. Rev.} \textbf{\bibinfo{volume}{D75}},
  \bibinfo{pages}{096003} (\bibinfo{year}{2007}{\natexlab{a}}),
  \eprint{hep-ph/0703042}.

\bibitem[{\citenamefont{Sun et~al.}(2007)\citenamefont{Sun, He, and
  Zhuang}}]{Sun:2007fc}
\bibinfo{author}{\bibfnamefont{G.-F.} \bibnamefont{Sun}},
  \bibinfo{author}{\bibfnamefont{L.}~\bibnamefont{He}}, \bibnamefont{and}
  \bibinfo{author}{\bibfnamefont{P.}~\bibnamefont{Zhuang}},
  \bibinfo{journal}{Phys. Rev.} \textbf{\bibinfo{volume}{D75}},
  \bibinfo{pages}{096004} (\bibinfo{year}{2007}), \eprint{hep-ph/0703159}.

\bibitem[{\citenamefont{Brauner}(2008)}]{Brauner:2008td}
\bibinfo{author}{\bibfnamefont{T.}~\bibnamefont{Brauner}},
  \bibinfo{journal}{Phys. Rev.} \textbf{\bibinfo{volume}{D77}},
  \bibinfo{pages}{096006} (\bibinfo{year}{2008}), \eprint{arXiv:0803.2422
  [hep-ph]}.

\bibitem[{\citenamefont{Chatterjee et~al.}(2008)\citenamefont{Chatterjee,
  Mishra, and Mishra}}]{Chatterjee:2008dr}
\bibinfo{author}{\bibfnamefont{B.}~\bibnamefont{Chatterjee}},
  \bibinfo{author}{\bibfnamefont{H.}~\bibnamefont{Mishra}}, \bibnamefont{and}
  \bibinfo{author}{\bibfnamefont{A.}~\bibnamefont{Mishra}}
  (\bibinfo{year}{2008}), \eprint{arXiv:0804.1051 [hep-ph]}.

\bibitem[{\citenamefont{He and Zhuang}(2007{\natexlab{b}})}]{He:2007yj}
\bibinfo{author}{\bibfnamefont{L.}~\bibnamefont{He}} \bibnamefont{and}
  \bibinfo{author}{\bibfnamefont{P.}~\bibnamefont{Zhuang}},
  \bibinfo{journal}{Phys. Rev.} \textbf{\bibinfo{volume}{D76}},
  \bibinfo{pages}{056003} (\bibinfo{year}{2007}{\natexlab{b}}),
  \eprint{arXiv:0705.1634 [hep-ph]}.

\bibitem[{\citenamefont{Cornwall et~al.}(1974)\citenamefont{Cornwall, Jackiw,
  and Tomboulis}}]{Cornwall:1974vz}
\bibinfo{author}{\bibfnamefont{J.~M.} \bibnamefont{Cornwall}},
  \bibinfo{author}{\bibfnamefont{R.}~\bibnamefont{Jackiw}}, \bibnamefont{and}
  \bibinfo{author}{\bibfnamefont{E.}~\bibnamefont{Tomboulis}},
  \bibinfo{journal}{Phys. Rev.} \textbf{\bibinfo{volume}{D10}},
  \bibinfo{pages}{2428} (\bibinfo{year}{1974}).

\bibitem[{\citenamefont{Deng et~al.}(2008)\citenamefont{Deng, Wang, and
  Wang}}]{Deng:2008ah}
\bibinfo{author}{\bibfnamefont{J.}~\bibnamefont{Deng}},
  \bibinfo{author}{\bibfnamefont{J.-c.} \bibnamefont{Wang}}, \bibnamefont{and}
  \bibinfo{author}{\bibfnamefont{Q.}~\bibnamefont{Wang}},
  \bibinfo{journal}{Phys. Rev.} \textbf{\bibinfo{volume}{D78}},
  \bibinfo{pages}{034014} (\bibinfo{year}{2008}), \eprint{arXiv:0803.4360
  [hep-ph]}.

\bibitem[{\citenamefont{Sa~de Melo et~al.}(1993)\citenamefont{Sa~de Melo,
  Randeria, and Engelbrecht}}]{Sa:1993sa}
\bibinfo{author}{\bibfnamefont{C.~A.~R.} \bibnamefont{Sa~de Melo}},
  \bibinfo{author}{\bibfnamefont{M.}~\bibnamefont{Randeria}}, \bibnamefont{and}
  \bibinfo{author}{\bibfnamefont{J.~R.} \bibnamefont{Engelbrecht}},
  \bibinfo{journal}{Phys. Rev. Lett.} \textbf{\bibinfo{volume}{71}},
  \bibinfo{pages}{3202} (\bibinfo{year}{1993}).

\bibitem[{\citenamefont{Diener et~al.}(2008)\citenamefont{Diener, Sensarma, and
  Randeria}}]{Diener:2007di}
\bibinfo{author}{\bibfnamefont{R.~B.} \bibnamefont{Diener}},
  \bibinfo{author}{\bibfnamefont{R.}~\bibnamefont{Sensarma}}, \bibnamefont{and}
  \bibinfo{author}{\bibfnamefont{M.}~\bibnamefont{Randeria}},
  \bibinfo{journal}{Phys. Rev.} \textbf{\bibinfo{volume}{A77}},
  \bibinfo{pages}{023626} (\bibinfo{year}{2008}), \eprint{arXiv:0709.2653
  [cond-mat.other]}.

\bibitem[{\citenamefont{Fetter and Walecka}(1971)}]{Fetter:1971fw}
\bibinfo{author}{\bibfnamefont{A.~L.} \bibnamefont{Fetter}} \bibnamefont{and}
  \bibinfo{author}{\bibfnamefont{J.~D.} \bibnamefont{Walecka}},
  \emph{\bibinfo{title}{Quantum theory of many-particle systems}},
  International series in pure and applied physics
  (\bibinfo{publisher}{McGraw--Hill}, \bibinfo{address}{New York},
  \bibinfo{year}{1971}).

\bibitem[{\citenamefont{Furnstahl and Hammer}(2002)}]{Furnstahl:2002gt}
\bibinfo{author}{\bibfnamefont{R.~J.} \bibnamefont{Furnstahl}}
  \bibnamefont{and} \bibinfo{author}{\bibfnamefont{H.~W.}
  \bibnamefont{Hammer}}, \bibinfo{journal}{Annals Phys.}
  \textbf{\bibinfo{volume}{302}}, \bibinfo{pages}{206} (\bibinfo{year}{2002}),
  \eprint{nucl-th/0208058}.

\bibitem[{\citenamefont{Thouless}(1960)}]{Thouless:1960th}
\bibinfo{author}{\bibfnamefont{D.~J.} \bibnamefont{Thouless}},
  \bibinfo{journal}{Ann. Phys. (N.Y.)} \textbf{\bibinfo{volume}{10}},
  \bibinfo{pages}{553} (\bibinfo{year}{1960}).

\bibitem[{\citenamefont{Nardulli}(2002)}]{Nardulli:2002ma}
\bibinfo{author}{\bibfnamefont{G.}~\bibnamefont{Nardulli}},
  \bibinfo{journal}{Riv. Nuovo Cim.} \textbf{\bibinfo{volume}{25N3}},
  \bibinfo{pages}{1} (\bibinfo{year}{2002}), \eprint{hep-ph/0202037}.

\bibitem[{\citenamefont{Bailin and Love}(1984)}]{Bailin:1984bm}
\bibinfo{author}{\bibfnamefont{D.}~\bibnamefont{Bailin}} \bibnamefont{and}
  \bibinfo{author}{\bibfnamefont{A.}~\bibnamefont{Love}},
  \bibinfo{journal}{Phys. Rept.} \textbf{\bibinfo{volume}{107}},
  \bibinfo{pages}{325} (\bibinfo{year}{1984}).

\bibitem[{\citenamefont{Pisarski and Rischke}(2000)}]{Pisarski:1999tv}
\bibinfo{author}{\bibfnamefont{R.~D.} \bibnamefont{Pisarski}} \bibnamefont{and}
  \bibinfo{author}{\bibfnamefont{D.~H.} \bibnamefont{Rischke}},
  \bibinfo{journal}{Phys. Rev.} \textbf{\bibinfo{volume}{D61}},
  \bibinfo{pages}{074017} (\bibinfo{year}{2000}), \eprint{nucl-th/9910056}.

\bibitem[{\citenamefont{Abuki and Kunihiro}(2006)}]{Abuki:2005ms}
\bibinfo{author}{\bibfnamefont{H.}~\bibnamefont{Abuki}} \bibnamefont{and}
  \bibinfo{author}{\bibfnamefont{T.}~\bibnamefont{Kunihiro}},
  \bibinfo{journal}{Nucl. Phys.} \textbf{\bibinfo{volume}{A768}},
  \bibinfo{pages}{118} (\bibinfo{year}{2006}), \eprint{hep-ph/0509172}.

\bibitem[{\citenamefont{Iida et~al.}(2004)\citenamefont{Iida, Matsuura,
  Tachibana, and Hatsuda}}]{Iida:2003cc}
\bibinfo{author}{\bibfnamefont{K.}~\bibnamefont{Iida}},
  \bibinfo{author}{\bibfnamefont{T.}~\bibnamefont{Matsuura}},
  \bibinfo{author}{\bibfnamefont{M.}~\bibnamefont{Tachibana}},
  \bibnamefont{and} \bibinfo{author}{\bibfnamefont{T.}~\bibnamefont{Hatsuda}},
  \bibinfo{journal}{Phys. Rev. Lett.} \textbf{\bibinfo{volume}{93}},
  \bibinfo{pages}{132001} (\bibinfo{year}{2004}), \eprint{hep-ph/0312363}.

\bibitem[{\citenamefont{Son}(1999)}]{Son:1998uk}
\bibinfo{author}{\bibfnamefont{D.~T.} \bibnamefont{Son}},
  \bibinfo{journal}{Phys. Rev.} \textbf{\bibinfo{volume}{D59}},
  \bibinfo{pages}{094019} (\bibinfo{year}{1999}), \eprint{hep-ph/9812287}.

\bibitem[{\citenamefont{Schaefer and Wilczek}(1999)}]{Schaefer:1999jg}
\bibinfo{author}{\bibfnamefont{T.}~\bibnamefont{Schaefer}} \bibnamefont{and}
  \bibinfo{author}{\bibfnamefont{F.}~\bibnamefont{Wilczek}},
  \bibinfo{journal}{Phys. Rev.} \textbf{\bibinfo{volume}{D60}},
  \bibinfo{pages}{114033} (\bibinfo{year}{1999}), \eprint{hep-ph/9906512}.

\bibitem[{\citenamefont{Giannakis et~al.}(2004)\citenamefont{Giannakis, Hou,
  Ren, and Rischke}}]{Giannakis:2004xt}
\bibinfo{author}{\bibfnamefont{I.}~\bibnamefont{Giannakis}},
  \bibinfo{author}{\bibfnamefont{D.-F.} \bibnamefont{Hou}},
  \bibinfo{author}{\bibfnamefont{H.-C.} \bibnamefont{Ren}}, \bibnamefont{and}
  \bibinfo{author}{\bibfnamefont{D.~H.} \bibnamefont{Rischke}},
  \bibinfo{journal}{Phys. Rev. Lett.} \textbf{\bibinfo{volume}{93}},
  \bibinfo{pages}{232301} (\bibinfo{year}{2004}), \eprint{hep-ph/0406031}.

\bibitem[{\citenamefont{Gorkov and Melik-Barkhudarov}(1961)}]{Gorkov:1961gm}
\bibinfo{author}{\bibfnamefont{L.~P.} \bibnamefont{Gorkov}} \bibnamefont{and}
  \bibinfo{author}{\bibfnamefont{T.~K.} \bibnamefont{Melik-Barkhudarov}},
  \bibinfo{journal}{Sov. Phys. JETP} \textbf{\bibinfo{volume}{13}},
  \bibinfo{pages}{1018} (\bibinfo{year}{1961}).

\bibitem[{\citenamefont{Furnstahl et~al.}(2007)\citenamefont{Furnstahl, Hammer,
  and Puglia}}]{Furnstahl:2006pa}
\bibinfo{author}{\bibfnamefont{R.~J.} \bibnamefont{Furnstahl}},
  \bibinfo{author}{\bibfnamefont{H.~W.} \bibnamefont{Hammer}},
  \bibnamefont{and} \bibinfo{author}{\bibfnamefont{S.~J.}
  \bibnamefont{Puglia}}, \bibinfo{journal}{Annals Phys.}
  \textbf{\bibinfo{volume}{322}}, \bibinfo{pages}{2703} (\bibinfo{year}{2007}),
  \eprint{nucl-th/0612086}.

\end{thebibliography}
\end{document}